\documentstyle[sprocl,]{article}
\input{psfig}
\def\NPB{{\em Nucl. Phys.} B}

\def\mco{\multicolumn}
\def\rr{\raggedright}

\def\be{\begin{equation}}
\def\ee{\end{equation}}
\def\bea{\begin{eqnarray}}
\def\eea{\end{eqnarray}}

\def\approxlt{\ifmmode \rlap{$<$}{}_{{}_{{}_{\textstyle\sim}}} \else%
$\rlap{$<$}{}_{{}_{{}_{\textstyle\sim}}}$\fi}
\def\approxgt{\ifmmode \rlap{$>$}{}_{{}_{{}_{\textstyle\sim}}} \else%
$\rlap{$>$}{}_{{}_{{}_{\textstyle\sim}}}$\fi}
\def\p{\ifmmode\pm\else$\pm$\fi}
\def\about{$\sim$}
\def\msun{\ifmmode \rm{M}_\odot \else M$_\odot$\fi}	
\def\mdot{\ifmmode \dot M \else $\dot M$\fi}	
\def\Lx{L$_x$}
\def\LEdd{L$_{Edd}$}
\def\mc{\multicolumn{2}{c}}
\def\rb{\raisebox{1.5ex}[0pt]}

\def\hide#1{}
\def\usec{$\mu$sec}
\begin{document}
\title{MILLISECOND OSCILLATIONS IN X-RAY BINARIES}
\author{M. VAN DER KLIS}
\address{Astronomical Institute ``Anton Pannekoek'', University of
Amsterdam\\Kruislaan 403, 1098 SJ Amsterdam, The Netherlands\\E-mail: 
michiel@astro.uva.nl}

\maketitle

\abstracts{The first millisecond X-ray variability phenomena from
accreting compact objects have recently been discovered with the Rossi
X-ray Timing Explorer.  Three new phenomena are observed from low-mass
X-ray binaries containing low-magnetic-field neutron stars:
millisecond pulsations, burst oscillations and kiloHertz
quasi-periodic oscillations. Models for these new phenomena involve
the neutron star spin, and orbital motion closely around the neutron
star and rely explicitly on our understanding of strong gravity and
dense matter. I review the observations of these new neutron-star
phenomena and possibly related ones in black-hole candidates, and
describe the attempts to use them to perform measurements of
fundamental physical interest in these systems.}

\section{Introduction}\label{sect:intro}

\vskip-12cm{\bf Submitted to the Annual Review of Astronomy and Astrophysics;
to appear September 2000}\vskip11.cm

The principal motivation for studying accreting neutron stars and
black holes is that these objects provide a unique window on the
physics of strong gravity and dense matter. One of the most basic
expressions of the compactness of these compact objects is the short
(0.1--1\,msec) dynamical time scale characterizing the motion of
matter under the influence of gravity near them. Millisecond
variability will naturally occur in the process of accretion of matter
onto a stellar-mass compact object, an insight that dates back to at
least Shvartsman (1971).  For example, hot clumps orbiting in an
accretion disk around black holes and neutron stars will cause
quasi-periodic variability on time scales of about a millisecond
(Sunyaev 1973). Accreting low-magnetic field neutron stars will reach
millisecond spin periods, which can be detected when asymmetric
emission patterns form on the star's surface during X-ray bursts
(Radhakrishnan \& Srinivasan 1984, Alpar et al. 1982; Shara 1982,
Livio \& Bath 1982, see also Joss 1978). These early expectations
have finally been verified in a series of discoveries with NASA's
Rossi X-Ray Timing Explorer (RXTE; Bradt et al. 1993) within 2.5 years
after launch on 30 December 1995.

In this review, I discuss these newly discovered phenomena and the
attempts to use them to perform measurements of fundamental physical
interest. I concentrate on millisecond oscillations, periodic and
quasi-periodic variations in X-ray flux with frequencies exceeding
10$^{2.5}$\,Hz, but I also discuss their relations to slower
variability and X-ray spectral properties. Millisecond oscillations
have so far been seen nearly exclusively from low-magnetic-field
neutron stars, so these will be the focus of this review, although I
shall compare their phenomenology to that of the black-hole-candidates
(\S\ref{sect:vuiltjes}-\ref{sect:models}).

Accreting neutron stars and black holes occur in X-ray binaries (e.g.,
Lewin et al. 1995a). In these systems matter is transferred from a
normal ('donor') star to a compact object.  Thermal X-rays powered by
the gravitational potential energy released are emitted by the inner
regions of the accretion flow and, if present, the neutron star
surface. For a compact object with a size of order 10$^1$\,km, 90\% of
the energy is released in the inner \about10$^2$\,km. It is with this
inner emitting region that we shall be mostly concerned here. Because
accreting low-magnetic-field neutron stars are mostly found in
low-mass X-ray binaries (in which the donor star has a mass of
$<$1\msun) these systems will be the ones we focus on.

The mass transfer usually occurs by way of an accretion disk around
the compact object. In the disk the matter moves in near-Keplerian
orbits, i.e., with an azimuthal velocity that is approximately
Keplerian and a radial velocity much smaller than this.  The disk has
a radius of 10$^{5-7}$\,km, depending on the binary separation.  The
geometry of the flow in the inner emitting regions is uncertain.  In
most models for accretion onto low-magnetic-field neutron stars (e.g.,
Miller et al. 1998a) at least part of the flow extends down into the
emitting region in the form of a Keplerian disk. It is terminated
either at the radius $R$ of the star itself, or at a radius $r_{in}$
somewhat larger than $R$, by for example the interaction with a weak
neutron-star magnetic field, radiation drag, or relativistic
effects. Within $r_{in}$ the flow is no longer Keplerian and may or may
not be disk-like. Both inside and outside $r_{in}$ matter may leave the
disk and either flow in more radially or be expelled. Particularly for
black holes advective flow solutions are discussed where the disk
terminates and the flow becomes more spherical at a much larger radius
(e.g., Narayan 1997).

Whatever the geometry, it is clear that as the characteristic
velocities near the compact object are of order
$(GM/R)^{1/2}\sim0.5c$, the dynamical time scale, the time scale for
the motion of matter through the emitting region, is short; $\tau_{dyn}
\equiv (r^3/GM)^{1/2}$$\sim$0.1\,ms for $r$=10\,km, and \about2\,ms
for $r$=100\,km near a 1.4\msun\ neutron star, and \about1\,ms at
100\,km from a 10\msun\ black hole. So, the significance of
millisecond X-ray variability from X-ray binaries is clear:
milliseconds is the natural time scale of the accretion process in the
X-ray emitting regions, and hence strong X-ray variability on such
time scales is nearly certainly caused by the motion of matter in
these regions. Orbital motion, neutron-star spin, disk- and
neutron-star oscillations are all expected to happen on these time
scales.

The inner flow is located in regions of spacetime where strong-field
general-relativity is required to describe the motion of matter. For
that reason one expects to detect strong-field general-relativistic
effects in these flows, such as for example the existence of a region
where no stable orbits are possible. The precise interactions between
the elementary particles in the interior of a neutron star which
determine the equation of state (EOS) of supra-nuclear-density matter
are not known. Therefore we can not confidently predict the radius of
a neutron star of given mass, or the maximum spin rate or mass of
neutron stars (e.g., Cook et al. 1994). So, by measuring these
macroscopic quantities one constrains the EOS and tests basic ideas
about the properties of elementary particles. In summary, the main
motivation for studying millisecond variations in X-ray binaries is
that their properties depend on untested, or even unknown, properties
of spacetime and matter.

Three different millisecond phenomena have now been observed in X-ray
binaries. Historically, the first to be discovered were the twin
kilohertz quasi-periodic oscillations (kHz QPOs), widely interpreted
now as due to orbital motion in the inner accretion flow. Then came
the burst oscillations, probably due to the spin of a layer in the
neutron star's atmosphere in near-corotation with the neutron star
itself. Finally RXTE detected the first true spin frequency of an
accreting low-magnetic field neutron star, the long-anticipated
accreting millisecond pulsar.

In this review, I first examine the millisecond pulsar
(\S\ref{sect:mspulses}), then the burst oscillations
(\S\ref{sect:burstosc}) and finally the kHz QPOs
(\S\ref{sect:khzqpos}). We will thus be venturing from the
(relatively) well-understood accreting pulsars via the less secure
regions of what happens in detail on a neutron star's surface during
the thermonuclear runaway that is an X-ray burst, into the mostly
uncharted territory of the innermost accretion flows around neutron
stars and black holes, which obviously is ``where the monsters are'',
but also where the greatest rewards wait. The possibly related
phenomena found in black-hole candidates (\S\ref{sect:bhc}) and at
lower frequencies (\S\ref{sect:slowqpo}) are discussed next, and in
\S\ref{sect:models} the kHz QPO models are summarized.

\section{Techniques}\label{sect:fft}

Most of the variability measurements discussed here rely on Fourier
analysis of X-ray count-rate time series with sub-millisecond time
resolution (van der Klis 1989b). A quasi-periodic oscillation (QPO) in
the time series stands out in the power spectrum (the square of the
Fourier transform) as a broad, usually Lorentzian peak (in
Fig.\,\ref{fig:twinpeaks} several of such peaks can be seen),
characterized by its frequency $\nu$ (``centroid frequency''), width
$\lambda$ (inversely proportional to the coherence time of the
oscillation) and strength (the peak's area is proportional to the
variance of the QPO signal). The variance is nearly always reported in
terms of the root-mean-square of the signal expressed as a fraction of
the count rate, the ``fractional rms amplitude'' $r$; the coherence
often in terms of a quality factor $Q=\nu/\lambda$. Conventionally, to
call a local maximum in a power spectrum a QPO peak one requires
$Q>2$. Time delays between signals simultaneously detected in
different energy bands are usually measured using cross-spectra (the
frequency-domain equivalent of the cross-correlation function; van der
Klis et al. 1987, Vaughan et al. 1994a, Nowak et al. 1998) and often
expressed in terms of a phase lag (time lag multiplied by frequency).

The signal-to-noise of a broad power-spectral feature is $n_\sigma =
{1\over 2} I_x r^2 (T/\lambda)^{1/2}$ (van der Klis 1989b, see van der
Klis 1998 for more details), where $I_x$ is the count rate and $T$ the
observing time (assumed $\gg1/\lambda$). Note that $n_\sigma$ is
proportional to the count rate and to the signal amplitude {\it
squared}, so that it is sufficient for the amplitude to drop by 50\%
for the signal-to-noise to go from, e.g., a whopping 6$\sigma$ to an
undetectable 1.5$\sigma$ -- i.e., if a power-spectral feature
``suddenly disappears'' it may have only decreased in amplitude by a
factor of two.

\section{Millisecond pulsations}\label{sect:mspulses}

An accreting millisecond pulsar in a low-mass X-ray binary has
sometimes been called the ``Holy Grail'' of X-ray astronomy. Its
discovery was anticipated for nearly 20 years, because magnetospheric
disk accretion theory as well as evolutionary ideas concerning the
genesis of millisecond {\it radio} pulsars strongly suggested that
such rapid spin frequencies must occur in accreting low-magnetic field
neutron stars (see Bhattacharya \& van den Heuvel 1991).  However, in
numerous searches of X-ray binary time series (e.g., Leahy et
al. 1983, Mereghetti \& Grindlay 1987, Wood et al. 1991, Vaughan et
al. 1994b) such rapid pulsars did not turn up.

\begin{figure}[htbp]
$$\psfig{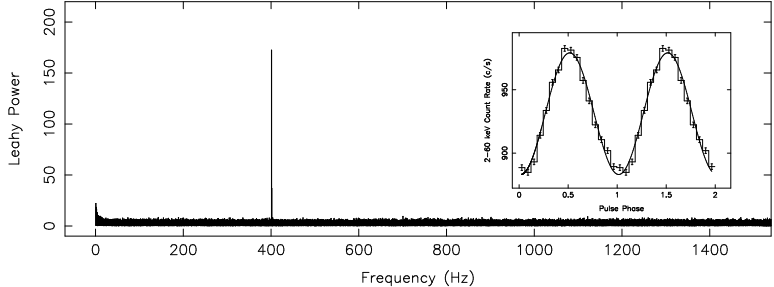}$$
\caption{The discovery power spectrum and pulse profile (inset) of the
first accreting millisecond X-ray pulsar. Note the low harmonic
content evident both from the absence of harmonics in the power
spectrum and the near-sinusoidal pulse profile. (after Wijnands \&
van der Klis 1998b)
\label{fig:1808fft}}
\end{figure}

More than two years after RXTE's launch, the first, and as of this
writing only accreting millisecond pulsar was finally discovered on
April 13, 1998 in the soft X-ray transient SAX\,J1808.4$-$3658
(Fig.~\ref{fig:1808fft}; Wijnands \& van der Klis 1998a,b). The pulse
frequency is 401\,Hz, so this is a 2.5 millisecond pulsar. The object
is nearly certainly the same as the transient that burst out at the
same position in September 1996 and gave the object its name (in 't
Zand et al. 1998). As this transient showed two type 1 X-ray bursts,
SAX\,J1808.4$-$3658 is also the first genuine bursting pulsar,
breaking the long-standing rule (e.g., Lewin \& Joss 1981) rule
that pulsations and type 1 X-ray bursts are mutually exclusive.

\begin{figure}[htbp]
\begin{center}
$$\psfig{figure=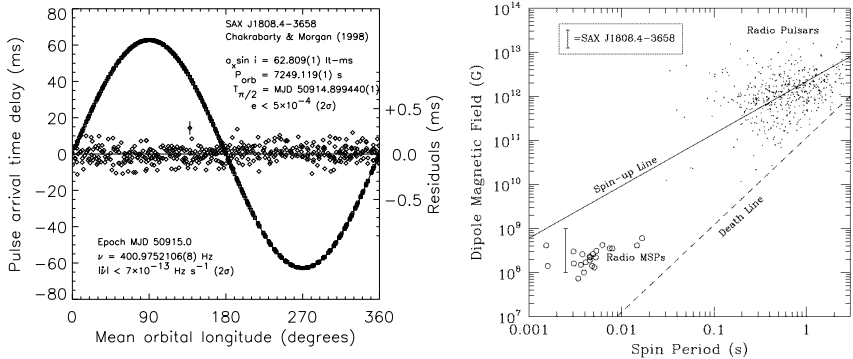,height=2in} 
\psfig{figure=psaltis_chaka_1808_ppdot.postscript,height=1.9in} $$
\end{center}
\caption{Left: the radial velocity curve and orbital elements of 
SAX\,J1808.4$-$3658. (Chakrabarty \& Morgan 1998b) Right: the
position of SAX\,J1808.4$-$3658 in the radio-pulsar period
vs. magnetic-field diagram. (Psaltis \& Chakrabarty 1999)
\label{fig:1808orbit}\label{fig:1808ppdot}}
\end{figure}

The orbital period of this pulsar is 2\,hrs
(Fig.\,\ref{fig:1808orbit}; Chakrabarty \& Morgan 1998a,b). With a
projected orbital radius $a\sin i$ of only 63 light {\it milli}seconds
and a mass function of 3.8\,10$^{-5}$\msun, the companion star is
either very low mass, or we are seeing the orbit nearly pole-on.  The
amplitude of the pulsations varied between 4 and 7\% and showed little
dependence on photon energy (Cui et al. 1998b). However, the
pulsations as measured at higher photon energies preceded those
measured in the 2-3\,keV band by a gradually increasing time interval
from 20\,\usec\ (near 3.5\,keV) up to 200\,\usec\ (between 10 and
25\,keV) (Cui et al. 1998b). These lags could be caused by Doppler
shifting of emission from the pulsar hot spots, higher-energy photons
being emitted earlier in the spin cycle as the spot approaches the
observer (Ford 1999).

An accreting magnetized neutron star spinning this fast must have a
weak magnetic field. If not, the radius of the magnetosphere $r_M$
would exceed the corotation radius, and matter corotating in the
magnetosphere would not be able to overcome the centrifugal barrier. A
simple estimate \hide{explain if space} leads to upper limits on $r_M$
of 31\,km, and on the surface field strength $B$ of
2--6\,10$^8$\,Gauss (Wijnands \& van der Klis 1998b). A similarly
simple estimate, involving in addition the requirement that for
pulsations to occur $r_M$ must be larger than the radius $R$ of the
neutron star, would set a strong constraint on the star's mass-radius
relation (Burderi \& King 1998\hide{explain if space}) and hence on
the EOS (see also Li et al. 1999a). However, the process of accretion
onto a neutron star with such a low $B$ is not identical to that in
classical, 10$^{12}$-Gauss accreting pulsars. In particular, the disk
model (used in calculating $r_M$) is different this close to the
neutron star, the disk-star boundary layer may be different, and
multipole components in the magnetic field become important.
Conceivably a classical magnetosphere does not even form and the
4--7\%-amplitude pulsation occurs due to milder effects of the
magnetic field on either the flow or the emission.  Psaltis \&
Chakrabarty (1999) discuss these issues and conclude that $B$ is
(1--10)\,10$^8$\,Gauss, which puts the source right among the other,
rotation-powered, millisecond pulsars (the msec radio pulsars;
Fig.\,\ref{fig:1808ppdot}). When the accretion shuts off sufficiently
for the radio pulsar mechanism to operate, the system will likely show
up as a radio pulsar. This should happen at the end of the system's
life as an X-ray binary, i.e., SAX\,J1808.4$-$3658 is indeed the
long-sought millisecond radio-pulsar progenitor, but might also occur
in between the transient outbursts (Wijnands \& van der Klis 1998b).
So far, radio observations have not detected the source in X-ray
quiescence (Gaensler et al. 1999).

It is not clear what makes the neutron star spin detectable in
SAX\,J1808.4$-$3658 and not (so far) in other low-mass X-ray binaries
of similar and often much higher flux. Perhaps a peculiar viewing
geometry (e.g., a very low inclination of the binary orbit) allows us
to see the pulsations only in this system, although possible X-ray and
optical modulations with binary phase make an inclination of zero
unlikely (Chakrabarty \& Morgan 1998b, Giles et al. 1999).

With a neutron star spin frequency that is certain and good estimates
of $r_M$ and $B$, SAX\,J1808.4$-$3658 can serve as a touchstone in
studies of low-mass X-ray binaries. Although no burst oscillations
(\S\ref{sect:burstosc}) or kHz QPOs (\S\ref{sect:khzqpos}) have been
detected from the source, their absence is consistent with what would
be expected from a standard LMXB in the same situation (Wijnands \&
van der Klis 1998c), and in more intensive observations during a next
transient outburst such phenomena could be detected. This would
strongly test the main assumptions underlying the models for these
phenomena. The X-ray spectral properties (Heindl \& Smith 1998,
Gilfanov et al. 1998) and the slower types of variability (Wijnands
\& van der Klis 1998c) of the source are very similar to those of
other LMXBs at low accretion rate suggesting that either the neutron
stars in those systems have similar $B$, or the presence of a small
magnetosphere does not affect spectral and slow-variability
characteristics.

\section{Burst oscillations}\label{sect:burstosc}

Type 1 X-ray bursts are thermonuclear runaways in the accreted matter
on a neutron-star surface (Lewin et al. 1995b for a review). When
density and temperature in the accumulated nuclear fuel approach the
ignition point, the matter ignites at one particular spot, from which
a nuclear burning front then propagates around the star (Bildsten
1998b for a review). This leads to a burst of X-ray emission with a
rise time of typically $<$1\,s, and a 10$^1$--10$^2$\,s exponential
decay due to cooling of the neutron-star atmosphere. The total amount
of energy emitted is 10$^{39-40}$\,erg. In some bursts the Eddington
critical luminosity is exceeded and atmospheric layers are 
lifted off the star's surface, leading to an increase in photospheric
radius of \about10$^1$--10$^2$\,km, followed by a gradual
recontraction. These bursts are called ``radius expansion bursts''.

In the initial phase, when the burning front is spreading, the energy
generation is inherently very anisotropic.  The occasional occurrence
of multiple bursts closely spaced in time indicates that not all
available fuel is burned up in each burst, suggesting that in some
bursts only part of the surface participates. Magnetic fields and
patchy burning (Bildsten 1995) could also lead to anisotropic emission
during X-ray bursts. Anisotropic emission from a spinning neutron star
leads to periodic or quasi-periodic observable phenomena, because due
to the stellar rotation the viewing geometry of the brighter regions
periodically varies (unless the pattern is symmetric around the
rotation axis). Searches for such periodic phenomena during X-ray
bursts were performed by various groups (Mason et al. 1980, Skinner et
al. 1982, Sadeh et al. 1982, Sadeh \& Livio 1982a,b, Murakami et
al. 1987, Schoelkopf \& Kelley 1991, Jongert \& van der Klis 1996),
but claims of detections remained unconfirmed.

\begin{figure}[htbp]
$$\psfig{figure=strohmayer_apj469L9_fig1.postscript,height=1.75in}\quad
\psfig{figure=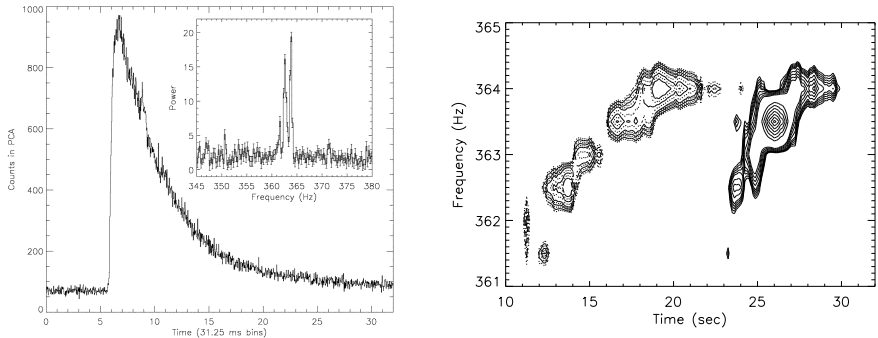,height=1.8in}$$
\caption{Left: a burst profile and its power spectrum (inset) showing
a drifting burst oscillation in 4U\,1728$-$34. (Strohmayer et
al. 1996c) Right: dynamic power spectra of burst oscillations in two
bursts separated by 1.6\,yr in 4U\,1728$-$34 showing near-identical
asymptotic frequencies. (Strohmayer et al. 1998b)
\label{fig:burstdrift} \label{fig:burstasymp}}
\end{figure}

The first incontestable type 1 burst oscillation was discovered with
RXTE in a burst that occurred on 1996 February 16 in the reliable
burst source 4U\,1728$-$34. An oscillation with a slightly drifting
frequency near 363\,Hz was evident in a power spectrum of 32\,s of
data starting just before the onset of the burst (Strohmayer et
al. 1996a,b,c; Fig.\,\ref{fig:burstdrift}). The oscilation frequency
increased from 362.5 to 363.9\,Hz in the course of \about10\,s.

Burst oscillations have now been detected in six (perhaps seven)
different sources (Table\,\ref{tab:burstosc}). They do not occur in
each burst, and some burst sources have not so far shown them at
all. Sometimes the oscillations are strong for less than a second
during the burst rise, then become weak or undetectable, and finally
occur for \about10\,s in the burst cooling tail. This can happen even
in radius expansion bursts, where after the photosphere has
recontracted the oscillations (re)appear (Smith et al. 1997,
Strohmayer et al. 1997a, 1998a). Some oscillations are seen only in
the burst tail and not in the rise (Smith et al. 1997).

\begin{table}[htbp]
\footnotesize
\caption{Burst oscillations\label{tab:burstosc}} 
\begin{center}
\footnotesize
\begin{tabular}{|lcccl|}
\hline
Source        & Minimum     &  \mc{Asymptotic}                 & References \\ 
              & observed    &  \mc{frequency}                  &            \\
              & frequency   &  \mc{range}                      &            \\
              & (Hz)        &  \mc{(Hz)}                       &            \\
\hline
4U\,1636$-$53 & 579.3   $^a$& 581.47\p0.01       & 581.75\p0.13& 1,2,3,4,5,15\\
4U\,1702$-$43 & 329.0       & 329.8\p0.1         & 330.55\p0.02& 6,14       \\
4U\,1728$-$34 & 362.1       & 363.94\p0.05       & 364.23\p0.05& 2,6,7,8    \\
KS\,1731$-$260& 523.9       & \mc{523.92\p0.05}                & 9,16        \\
MXB\,1743$-$29$^b$& 588.9   & \mc{589.80\p0.07}                & 10,17       \\
Aql\,X-1      & 547.8       & \mc{548.9}                       & 11,12,18    \\
Rapid Burster & 154.9; 306.6$^c$&                &             & 13         \\
\hline
\end{tabular}
\end{center}
$^a$ A weak subharmonic exists near 290 Hz (Miller 1999a). $^b$ Source
identification uncertain. $^c$ Marginal detections.  References: [1]
Strohmayer et al. 1998a [2] Strohmayer et al. 1998b [3] Strohmayer 1999
[4] Miller 1999a [5] Miller 1999b [6] Strohmayer \& Markwardt 1999 [7]
Strohmayer et al. 1996b,c [8] Strohmayer et al. 1997b [9] Smith et
al. 1997 [10] Strohmayer et al. 1997a [11] Zhang et al. 1998c [12]
Ford 1999 [13] Fox et al. 1999 [14] Markwardt et al. 1999a [15] Zhang
et al. 1997a [16] Morgan and Smith 1996 [17] Strohmayer et al. 1996d
[18] Yu et al. 1999
\end{table}

\begin{figure}[htbp]
$$\psfig{figure=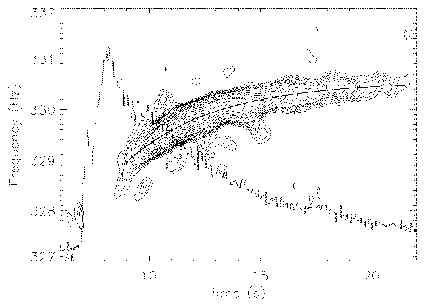,height=1.7in}
\psfig{figure=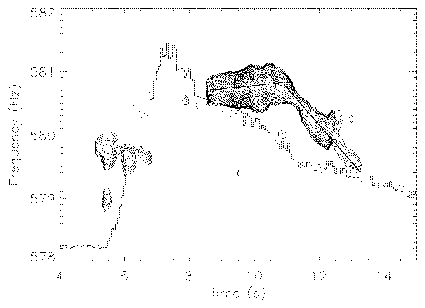,height=1.7in}$$
\caption{Dynamic power spectra of burst oscillations (contours)
overlaid on burst flux profiles (traces). Left: strong burst
oscillation in 4U\,1702$-$43 showing the usual asymptotic increase in
frequency. (Strohmayer \& Markwardt 1999) Right: oscillation in
4U\,1636$-$53 exhibiting a drop in frequency in the burst
tail. (Strohmayer 1999) \hide{this figure can be taken out in a
squeeze}
\label{fig:burstwrongdrift}}
\end{figure}

Usually, the frequency increases by 1--2\,Hz during the burst tail,
converging to an ``asymptotic frequency'' which in a given source
tends to be stable ((Strohmayer et al. 1998c;
Fig.\,\ref{fig:burstasymp}), with differences from burst to burst of
\about0.1\% (Table\,\ref{tab:burstosc}). This is of the order of what 
would be expected from binary orbital Doppler shifts and strengthens
an interpretation in terms of the neutron star spin. Perhaps the
orbital radial velocity curve can be detected in the asymptotic
frequencies, but it remains to be seen if the asymptotic frequencies
are intrinsically stable enough for this.  Exceptions from the usual
frequency evolution pattern do occur.  Oscillations with no evidence
for frequency evolution were observed in KS\,1731$-$26 (Smith et
al. 1997) and 4U\,1743$-$29 (Strohmayer et al. 1997a), and in
4U\,1636$-$53 a {\it decrease} in frequency was seen in a burst tail
(Miller 1999b, Strohmayer 1999; Fig.\,\ref{fig:burstwrongdrift}).
\hide{in the 1743 a reference is given to no freq evol bursts in
4U\,1728$-$34 attributed to Strohmayer et al. (1997) but this never
materialized}

In a widely (but not universally; \S\ref{sect:rpm}) accepted scenario,
the burst oscillations arise due to a hot spot or spots in an
atmospheric layer of the neutron star rotating slightly slower than
the star itself because it expanded by 5--50\,m in the X-ray burst but
conserved its angular momentum (Strohmayer et al. 1997a and references
therein, Bildsten 1998b, Strohmayer 1999, Strohmayer \& Markwardt
1999, Miller 1999b). The frequency drifts are caused by spin-up of the
atmosphere as it recontracts in the burst decay. The asymptotic
frequency corresponds to a fully recontracted atmosphere and is
closest to the true neutron star spin frequency. From this scenario
one expects a frequency drop during the burst rise, but no good
evidence has been found for this as yet (Strohmayer 1999). The case of
a frequency drop in the burst tail (Fig.\,\ref{fig:burstwrongdrift})
is explained by invoking additional thermonuclear energy input late in
the burst, which also affects the burst profile (Strohmayer 1999).

If the oscillations are due to a stable pattern in the spinning layer,
then it should be possible to describe them as a frequency-modulated,
strictly coherent signal. By applying a simple exponential model to
the frequency drifts, it is possible to establish coherences of up to
Q\about4000 (Strohmayer \& Markwardt 1999, see also Zhang et
al. 1998c, Smith et al. 1997, Miller 1999a,b). However, this is still
\about20\% less than a fully coherent signal of this frequency and
duration, i.e., it has not yet been possible to count the exact number of
cycles in the way this can be done in a pulsar.  Possibly, exact
coherence recovery is feasible for these signals, but current
signal-to-noise limits prevent to accomplish this as the exact
frequency drift ephemeris can not be found.

The harmonic content of the oscillations is low. In 4U\,1636$-$53 it
is just possible, by combining data from the early stages of several
bursts, to detect a frequency near 290\,Hz, half the dominant one
(Miller 1999a,b), suggesting that \about290\,Hz, not \about580\,Hz is
the true spin frequency and that two antipodal hot spots produce the
burst oscillation in this source (whose kHz peak separation is
\about250\,Hz, \S\ref{sect:khzqpos}). No harmonics or ``subharmonics''
have been seen in other sources, or in the burst tails of any source
(Strohmayer \& Markwardt 1999), with the possible exception of the
marginal detections in the Rapid Burster (Fox et al. 1999).

The oscillation amplitudes range from \about50\% (rms) of the burst
flux early in some bursts to between 2 and 20\% (rms) in the tail
(references see Table\,\ref{tab:burstosc}). (Note, that sometimes
sinusoidal amplitudes are reported, which are a factor $\sqrt2$ larger
than rms amplitudes, and that amplitudes are expressed as a fraction
of {\it burst} flux, total flux minus the persistent flux before the
burst. Early in the rise, when burst flux is low compared to total
flux, the measured amplitude is multiplied by a large factor to
convert it to burst flux fraction.) In KS\,1731$-$260 (Smith et
al. 1997) and 4U\,1743$-$29 (Strohmayer et al. 1997a) the photon
energy dependence of the oscillations was measured. Above 7 or 8\,keV
the amplitude was 9--18\% while below that energy it was undetectable
at $<$2--4\%. 4U\,1636$-$53 shows a slight variation in spectral
hardness as a function of oscillation phase (Strohmayer et
al. 1998). In Aql\,X-1 photons below 5.7\,keV lag those at higher
energies by roughly 0.3\,msec, which may be caused by Doppler shifts
(Ford 1999). 

If the burst oscillations are due to hot spots on the surface, then
their amplitude constrains the ``compactness'' of the neutron star,
defined as $R_G/R$ where $R$ is the star's radius and $R_G=GM/c^2$ its
gravitational radius (Strohmayer et al. 1997b, 1998a, Miller \& Lamb
1998). The more compact the star, the lower the oscillation amplitude,
as gravitational light bending increasingly blurs the beam. In
particular when the oscillations are caused by two antipodal hot spots
(cf. 4U\,1636$-$53, above, and possibly other sources,
\S\ref{sect:deltanu}), and the amplitudes are high, the constraints
are strong. The exact bounds on the compactness depend on the emission
characteristics of the spots, and no final conclusions have been
reached yet.

Modeling of the spectral and amplitude evolution of the oscillations
through the burst in terms of an expanding, cooling hot spot has been
succesful (Strohmayer 1997b). However, several issues with respect to
this attractively simple interpretation have yet to be resolved. The
presence of {\it two} burning sites as required by the description in
terms of two antipodal hot spots could be related to fuel accumulation
at the magnetic poles (Miller 1999a), but their simultaneous ignition
seems not easy to accomplish, and obviously, in view of the frequency
drifts, these sites must decouple from the magnetic field after
ignition. The hot spots must survive the strong shear in layers that,
from the observed phase drifts, must revolve around the star several
times during the lifetime of the spots, and they must even survive
through photospheric radius expansion by factors of at least several
during radius-expansion bursts (which probably implies the roots of
the hot spots are below the photospheric layers).

The resolution of these issues ties in with questions such as why only
some burst sources show the oscillations, and why only some bursts
exhibit them. This is hard to explain in a magnetic-pole accumulation
scheme as the viewing geometry of the poles remains the same from
burst to burst. Studies of the relation between the characteristics of
the bursts and the surrounding persistent emission and the presence
and character of the burst oscillations could help to shed light on
these various questions.

\section{Kilohertz quasi-periodic oscillations}\label{sect:khzqpos}

\begin{figure}[htbp]
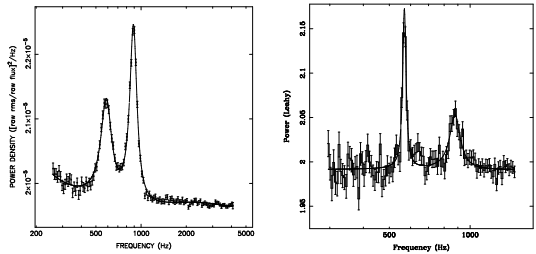

$$\psfig{figure=apjl_xte_sco_horne_fig1.postscript,height=1.3in}\quad
\psfig{figure=1608_mendez_apj505L23_fig2.postscript,height=1.3in}$$
\caption{Twin kHz peaks in Sco\,X-1 (left; van der Klis et al. 1997b)
and 4U\,1608$-$52 (right; M\'endez et al. 1998b).
\label{fig:twinpeaks}}
\end{figure}

The kilohertz quasi-periodic oscillations (kHz QPOs) were discovered
at NASA's Goddard Space Flight Center in February 1996, just two
months after RXTE was launched (in Sco\,X-1: van der Klis et
al. 1996a,b,c, and 4U\,1728$-$34: Strohmayer et al. 1996a,b,c; see van
der Klis 1998 for a historical account). Two simultaneous
quasi-periodic oscillation peaks (``twin peaks'') in the 300--1300\,Hz
range and roughly 300\,Hz apart (Fig.\,\ref{fig:twinpeaks}) occur in
the power spectra of low-mass X-ray binaries containing
low-magnetic-field neutron stars of widely different X-ray luminosity
\Lx. The frequency of both peaks usually increases with X-ray flux
(\S\ref{sect:parlines}). In 4U\,1728$-$34 the separation frequency of
the two kHz peaks is close to $\nu_{burst}$ (\S\ref{sect:burstosc};
Strohmayer et al. 1996b,c). This commensurability of frequencies
provides a powerful argument for a beat-frequency interpretation
(\S\ref{sect:formulas},
\ref{sect:deltanu}, \ref{sect:spm}).

\subsection{Orbital and beat frequencies}\label{sect:formulas}

\begin{figure}[htbp]
$$\psfig{figure=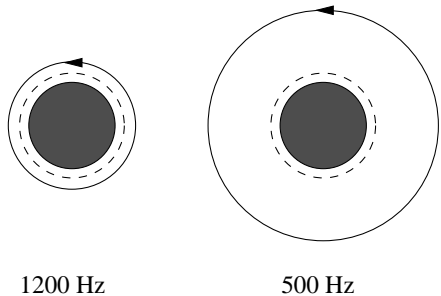,height=1.5in,angle=-90}$$
\caption{A 10-km radius, 1.4\msun\ neutron star with the
corresponding innermost circular stable orbit (ISCO; dashed circles)
and orbits (drawn circles) corresponding to orbital frequencies of
1200 and 500\,Hz, drawn to scale.
\label{fig:circles}}
\end{figure}

Orbital motion around a neutron star occurs at a frequency of
$$\nu_{orb} =\left(GM\over4\pi^2r_{orb}^3\right)^{1/2} \approx
1200\,\hbox{Hz}\,\left(r_{orb}\over15\,\hbox{km}\right)^{-3/2}m_{1.4}^{1/2},$$
and the corresponding orbital radius is
$$r_{orb} = \left(GM\over4\pi^2\nu_{orb}^2\right)^{1/3} \approx
15\,\hbox{km}\,\left(\nu_{orb}\over1200\,\hbox{Hz}\right)^{-2/3}m_{1.4}^{1/3},$$
where $m_{1.4}$ the star's mass in units of 1.4\msun\
(Fig.\,\ref{fig:circles}). In general relativity, no stable orbital
motion is possible within the innermost stable circular orbit (ISCO),
$R_{ISCO} = 6GM/c^2 \approx 12.5m_{1.4}\,\hbox{km}\,$. The frequency
of orbital motion at the ISCO, the highest possible stable orbital
frequency, is $\nu_{ISCO}
\approx (1580/m_{1.4})\,\hbox{Hz}$.

These expressions are valid for a Schwarzschild geometry, i.e.,
outside a non-rotating spherically symmetric neutron star (or black
hole). Corrections to first order in $j=cJ/GM^2$, where $J$ is the
neutron-star angular momentum have been given by, e.g., Miller et
al. (1998a) and can be several 10\%. For more precise calculations see
Morsink \& Stella (1999).

In spin-orbit beat-frequency models some mechanism produces an {\it
interaction} of $\nu_{orb}$ at some preferred radius in the accretion
disk with the neutron star spin frequency $\nu_s$, so that a beat
signal is seen at the frequency $\nu_{beat} = \nu_{orb}-\nu_s$.  As
$\nu_{beat}$ is the frequency at which a given particle orbiting in
the disk overtakes a given point on the spinning star, it is the
natural disk/star interaction frequency. In such a rotational
interaction (with spin and orbital motion in the same sense) no signal
is produced at the sum frequency $\nu_{orb}+\nu_s$ (it is a
``single-sideband'' interaction).

\subsection{Early interpretations}\label{sect:early}

It was immediately realized that the observed high frequencies of kHz
QPOs could arise in orbital motion of accreting matter very closely
around the neutron star, or in a beat between such orbital motion and
the neutron-star spin (van der Klis et al. 1996a, Strohmayer et
al. 1996a; some of the proposals to look for such rapid QPOs with RXTE
had in fact anticipated this).  A magnetospheric spin-orbit
beat-frequency model (Alpar \& Shaham 1985, Lamb et al. 1985) was
already in use in LMXBs for slower QPO phenomena
(\S\ref{sect:slowqpo}) so when three commensurable frequencies were
found in 4U\,1728$-$34, a beat-frequency interpretation was
immediately proposed (Strohmayer et al. 1996c).  Let us call $\nu_2$
the frequency of the higher-frequency (the ``upper'') peak and $\nu_1$
that of the lower-frequency kHz peak (the ``lower peak'').  Then the
beat-frequency interpretation asserts that $\nu_2$ is $\nu_{orb}$ at
some preferred radius in the disk, and $\nu_1$ is the beat frequency
between $\nu_2$ and $\nu_s$, so $\nu_1 = \nu_{beat} = \nu_{orb} -
\nu_s \approx \nu_2 - \nu_{burst}$, where the approximate equality 
follows from $\nu_{burst}\approx\nu_s$ (\S\ref{sect:burstosc}). That
only a single sideband is observed is a strong argument for a
rotational interaction (\S\ref{sect:formulas}).  Later this was worked
out in detail in the form of the ``sonic point beat-frequency model''
by Miller et al. (1996, 1998a; \S\ref{sect:spm}), where the preferred
radius is the sonic radius (essentially, the inner edge of the
Keplerian disk).

The beat-frequency interpretation implies that the observed kHz QPO
peak separation $\Delta\nu=\nu_2-\nu_1$ should be equal to the neutron
star spin frequency $\nu_s$, and should therefore be constant and
nearly equal to $\nu_{burst}$. As we shall see in
\S\ref{sect:deltanu}, it turned out that $\Delta\nu$ is not actually
exactly constant nor precisely equal to $\nu_{burst}$ (in
\S\ref{sect:models}  we examine how a beat-frequency interpretation 
could deal with this), and this triggered the development of other
models for the kHz QPOs. Stella \& Vietri (1998) noted that the
frequency $\nu_{LF}$ of low-frequency QPOs (\S\ref{sect:slowqpo}) in
the 15--60\,Hz range that had been known in the Z sources since the
1980's (cf. van der Klis 1995), and that were being discovered with
RXTE in the atoll sources as well is approximately proportional to
${\nu_2}^2$. This triggered a series of papers (Stella \& Vietri
1998, 1999, Stella et al. 1999b) together describing what is now
called the ``relativistic precession model'', where $\nu_2$ is the
orbital frequency at some radius in the disk and $\nu_1$ and
$\nu_{LF}$ are frequencies of general-relativistic precession modes of
a free particle orbit at that radius (\S\ref{sect:rpm}). For a further
discussion of kHz QPO models see
\S\ref{sect:models}.  With the exception of the photon bubble model
(Klein et al. 1996b; \S\ref{sect:pbm}), all models are based on the
interpretation that one of the kHz QPO frequencies is an orbital
frequency in the disk.

\subsection{Dependence on source state and type}\label{sect:zatol}

Twenty sources have now\footnote{1999 October 15} shown kHz
QPOs. Sometimes only one peak is detectable, but 18 of these sources
have shown two simultaneous kHz peaks; the exceptions with only a
single peak are the little-studied XTE\,J1723$-$376 and
EXO\,0748$-$676.  Tables~\ref{tab:z} and \ref{tab:atoll} summarize the
results.  There is a remarkable similarity in QPO frequencies and peak
separations across a great variety of sources.

\begin{table}[htbp]
\caption{Observed frequencies of kilohertz QPOs in Z sources} 
\label{tab:z}
\scriptsize
\begin{center}
\begin{tabular}{|lccccl|}
\hline
Source &$\nu_1$&$\nu_2$&$\Delta\nu$&$\nu_{burst}$   & References\\ 
       & (Hz)  & (Hz)  & (Hz)      & (Hz)           &\\
\hline 
Sco\,X-1        & 565    & 870     & 307$\pm$5     && Van der Klis et al. 1996a,b,c,1997b \\
                & 845    & 1080    & 237$\pm$5     && \\
                &        & 1130    &               && \\
\hline					       
GX\,5$-$1       & 215    & 505     &               && Van der Klis et al. 1996e \\
                & 660    & 890     &\rb{298$\pm$11}&& Wijnands et al. 1998c \\
                & 700    &         &               && \\
\hline					       
GX\,17+2        &        & 645     &               && Van der Klis et al. 1997a \\
                & 480    & 785     &               && Wijnands et al. 1997b \\
                & 780    & 1080    &\rb{294$\pm$8} && \\
\hline
Cyg\,X-2        &        & 730     &               && Wijnands et al. 1998a \\
                & 530    & 855     &               && \\
                & 660    & 1005    &\rb{346$\pm$29}&& \\
\hline
GX\,340+0       & 200    & 535     &               && Jonker et al. 1998, 1999c \\
                & 565    & 840     &\rb{339$\pm$8} && \\
                & 625    &         &               && \\
\hline
GX\,349+2       & 710    & 980     & 266$\pm$13    && Zhang et al. 1998a \\
                & \mc{1020$^a$}    &               && Kuulkers \& van der Klis 1998 \\
\hline
\end{tabular}
\end{center}
\vbox{Values for $\nu_1$ and $\nu_2$ were rounded to the nearest 5,
for $\nu_{burst}$ to the nearest 1\,Hz.  Entries in one column not
separated by a horizontal line indicate ranges over which the
frequency was observed or inferred to vary; ranges from different
observations were combined assuming the $\nu_1$, $\nu_2$ relation in
each source is reproducible (no evidence to the contrary exists).
Entries in one uninterrupted row refer to simultaneous data (except
for $\nu_{burst}$ values).  Values of $\Delta\nu$ straddling two rows,
or adjacent to a vertical line refer to measurements made over the
range of frequencies indicated. Note: $^a$ Marginal detection.}
\end{table}

\begin{table}[htbp]
\caption{Observed frequencies of kilohertz QPOs in atoll sources} 
\label{tab:atoll}
\tiny
\begin{center}
\begin{tabular}{|lccccl|}
\hline
Source          &$\nu_1$ &$\nu_2$  &$\Delta\nu$ &$\nu_{burst}$& References\\ 
                & (Hz)   & (Hz)    & (Hz)       & (Hz)$^d$    &\\
\hline					       
4U\,0614+09     &        & 450     &            &             & \smash{\hbox{\hsize=4.4cm\rr\vtop{Ford et 
                                                                al. 1996, 1997a,b; Van der Klis et 
                                                                al. 1996d;  M\'endez et al. 1997; Vaughan 
                                                                et al. 1997,1998; Kaaret et al. 1998; van 
                                                                Straaten et al. 1999}}} \\
                & 418    & 765     &            &             & \\
                & 825    & 1160    &\rb{312$\pm$2}&           & \\
                &        & 1215    &            &             & \\
                &        & 1330    &            &             & \\ 
\hline
EXO\,0748$-$676 & \mc{695}         &            &             & J. Homan 1999, in prep. \\ 
\hline
4U\,1608$-$52   & 415    &         &            &             & \smash{\hbox{\hsize=4.4cm\rr\vtop{Van 
                                                                Paradijs et al. 1996; Berger et al. 1996; 
                                                                Yu et al. 1997; Kaaret et al. 1998; Vaughan
                                                                et al. 1997,1998; M\'endez et al. 1998a,b,
                                                                1999; M\'endez 1999; Markwardt et al. 
                                                                1999b}}}\\            
                & 440    & 765     & 325$\pm$7  &             & \\ 
                & 475    & 800     & 326$\pm$3  &             & \\
                & 865    & 1090$^b$& 225$\pm$12$^b$&          & \\ 
                & 895    &         &            &             & \\  
\hline							        
4U\,1636$-$53   & 830    &         & \mco{1}{|}{} &           & \smash{\hbox{\hsize=4.4cm\rr\vtop{Zhang et
                                                                al. 1996a,b,1997a; Van der Klis et al. 1996d; 
                                                                Vaughan et al. 1997,1998; Zhang 1997; 
	                                                        Wijnands et al. 1997a; M\'endez et al. 
                                                                1998c; M\'endez 1999; Markwardt et al. 
                                                                1999b;
Kaaret et al. 1999a}}} \\
                & 900    & 1150    &\mco{1}{|}{}   &            & \\ 
                & 950    & 1190    &\mco{1}{|c}{251$\pm$4$^b$}&291,582& \\ 
                &        & 1230    &\mco{1}{|}{}   &            & \\ 
                & 1070   &         &\mco{1}{|}{}   &            & \\ 
\hline
4U\,1702$-$43   & 625    &         &             &             & Markwardt et al. 1999a,b \\
                & 655    & 1000$^b$&344$\pm$7$^b$&            & \\
                & 700    & 1040$^b$&337$\pm$7$^b$&  330       & \\ 
                & 770    & 1085$^b$&315$\pm$11$^b$&           & \\ 
                & 902    &         &            &             & \\  
\hline
4U\,1705$-$44   & 775    & 1075$^a$& 298$\pm$11 &             & Ford et al. 1998a \\ 
                & 870    &         &            &             & \\ 
\hline
XTE\,J1723$-$376& \mc{815}         &            &             & Marshall \& Markwardt 1999 \\
\hline
4U\,1728$-$34   &        &  325    &            &             & \smash{\hbox{\hsize=4.4cm\rr\vtop{Strohmayer
                                                                et al. 1996a,b,c; Ford \& van der Klis 
                                                                1998; M\'endez \& van der Klis 1999; 
                                                                M\'endez 1999; Markwardt et al. 1999b; 
                                                                di Salvo et al. 1999}}} \\
                & 510    &  845    &            &             & \\
                & 875    & 1160    &\mco{1}{|c}{349$\pm$2$^c$}&364& \\
                & 920    &         &\mco{1}{|c}{279$\pm$12$^c$}&& \\
\hline					       
KS\,1731$-$260  & 900    & 1160    & 260$\pm$10 & 524         & Wijnands \& van der Klis 1997 \\
                &        & 1205    &            &             & \\
                &        &         &            &             & \\
\hline
4U\,1735$-$44   & 630    & 980     & 341$\pm$7  &             & \smash{\hbox{\hsize=4.4cm\rr\vtop{Wijnands 
                                                                et al. 1996a, 1998b; Ford et al. 1998b}}} \\
                & 730    & 1025    & 296$\pm$12 &             & \\
                & 900$^a$& 1150    &  249$\pm$15&             & \\ 
                &        & 1160    &            &             & \\
\hline
4U\,1820$-$30   &        & 655     &            &             & \smash{\hbox{\hsize=4.4cm\rr\vtop{Smale et
                                                                al. 1996, 1997; Zhang et al. 1998b; Kaaret
                                                                et al. 1999b; Bloser et al. 1999}}} \\
                & 500    & 860     & 358$\pm$42 &             & \\
                & 795    & 1075    & 278$\pm$11 &             & \\
                &        & 1100    &            &             & \\
\hline
Aql\,X-1        & 670    &         &            &             & \smash{\hbox{\hsize=4.4cm\rr\vtop{Zhang et
                                                                al. 1998c; Cui et al. 1998a; Yu et al. 1999;
                                                                Reig
et al. 1999; M. M\'endez et al. 1999 in prep.}}}\\
                & 930    &\rb{1040$^a$}&\rb{241$\pm$9$^a$}&549& \\
                &        &         &            &             & \\
\hline
4U\,1915-05     &        & 820     &            &             & \smash{\hbox{\hsize=4.4cm\rr\vtop{Barret et 
                                                                al. 1997,1998; Boirin et al. 1999}}} \\
                & 515    &         &            &             & \\
                & 560    & 925     & \mco{1}{|c}{}&           & \\
                & 655    & 1005    & \mco{1}{|c}{348$\pm$11}& & \\
                & 705$^a$& 1055    & \mco{1}{|c}{}&           & \\
                & 880    &         &            &             & \\
                &        & 1265$^a$&            &             & \\
\hline
XTE\,J2123$-$058& 845    & 1100    &255$\pm$14  &             & \smash{\hbox{\hsize=4.4cm\rr\vtop{Homan et
                                                                al. 1998b,1999a; Tomsick et al. 1999}}} \\
                & 855    & 1130    &276$\pm$9   &             & \\
                & 870$^a$& 1140    &270$\pm$5$^a$&            & \\
\hline
\end{tabular}
\end{center}
\vbox{Caption: see Table~\ref{tab:z}. Notes: $^a$ Marginal detection.  $^b$
Shift and add detection method, cf. M\'endez et al. (1998a).  $^c$ See
Fig.\,\ref{fig:deltanuchanges}. $^d$ For burst oscillation references
see Table\,\ref{tab:burstosc}.}
\end{table}

\begin{figure}[htbp]
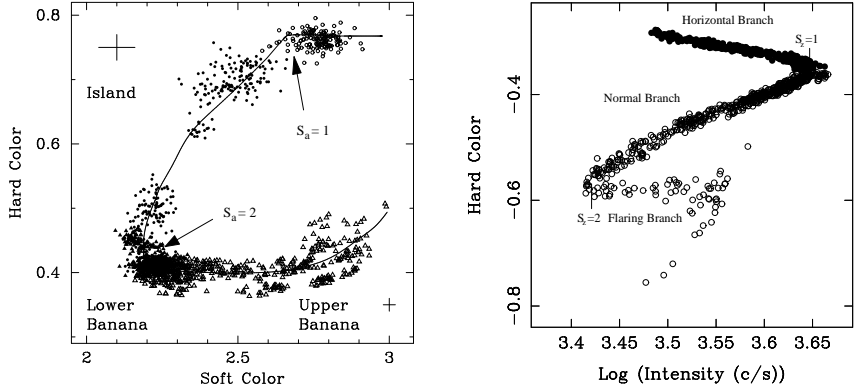

$$\psfig{figure=mendez_1608_apjlxxx_fig1_cd.ps,height=2.in,angle=-90}\qquad
\psfig{figure=340p0_jonker.postscript,height=2.in}$$
\caption{Left: X-ray color-color diagram of the atoll source
4U\,1608$-$52. (M\'endez et al. 1999)  Right: X-ray hardness
vs. intensity diagram of the Z source GX\,340+0. (after Jonker et
al. 1999c) Values of curve-length parameters S$_z$ and S$_a$ and
conventional branch names are indicated. Mass accretion rate is
inferred to increase in the sense of increasing S$_a$ and S$_z$. X-ray
color is the log of a count rate ratio (3.5--6.4/2.0--3.5 and
9.7--16.0/6.4--9.7 keV for soft and hard color respectively);
intensity is in the 2--16\,keV band. kHz QPO detections are indicated
with filled symbols. \label{fig:1608cd}
\label{fig:340hid}}
\end{figure}

However, at a more detailed level there turn out to be differences
between the different source types. The two main types are the Z
sources and the atoll sources (Hasinger \& van der Klis 1989, see van
der Klis 1989a, 1995a,b for reviews). Z sources are named after the
roughly Z-shaped tracks they trace out in X-ray color-color and
hardness-intensity diagrams on a time scale of hours to days
(Fig.\,\ref{fig:340hid}).  They are the most luminous LMXBs, with
X-ray luminosity \Lx\ near the Eddington luminosity \LEdd. Atoll
sources produce tracks roughly like a wide U or C (e.g.,
Fig.\,\ref{fig:1608cd}) which are somewhat reminiscent of a
geographical map of an atoll because in the left limb of the U the
motion through the diagram becomes slow, so that the track is usually
broken up by observational windowing into ``islands''. The bottom and
right-hand parts of the U are traced out in the form of a curved
``banana'' branch on a time scale of hours to a day. In a given
source, the islands correspond to lower flux levels than the banana
branch. Most atoll sources are in the 0.01--0.2 \LEdd\ range; a group
of four bright ones (GX\,3+1, GX\,9+1, GX\,13+1 and GX\,9+9) that is
nearly always in the banana branch is more luminous than this, perhaps
0.2--0.5\,\LEdd\ (the distances are uncertain). Most timing and
spectral characteristics of these sources depend in a simple way on
position along the Z or atoll track. So, the phenomenology is
essentially one-dimensional. A single quantity, usually referred to as
``inferred accretion rate'', varying on time scales of hours to days
on the Z track and the banana branch, and more slowly in the island
state must govern most of the phenomenology (but see \S\ref{sect:Lx}).

In all Z sources and in 4U\,1728$-$34 the kHz QPOs are seen down to
the lowest inferred \mdot\ levels these sources reach. The QPOs always
become undetectable at the highest \mdot\ levels. In the atoll
sources, where the count rates are higher at higher inferred \mdot,
this can not be a sensitivity effect. In most atoll sources, the QPOs
are seen in the part of the banana branch closest to the islands,
i.e., near the lower left corner of the U (Fig.~\ref{fig:1608cd});
that they are often not detected in the island state may be related to
low sensitivity at the low count rates there, but in one island in
4U\,0614+09 the undetected lower kHz peak is really much weaker than
at higher inferred \mdot\ (M\'endez et al. 1997). No kHz QPOs have
been seen in the four bright atoll sources (Wijnands et al. 1998d,
Strohmayer 1998, Homan et al. 1998a), perhaps because they do not
usually reach this low part of the banana branch, and in several faint
LMXBs, probably also atoll sources in the island state
(SAX\,J1808.4$-$3658; Wijnands \& van der Klis 1998c and
\S\ref{sect:mspulses}, XTE\,J1806$-$246; Wijnands \& van der Klis
1999b, SLX\,1735$-$269; Wijnands \& van der Klis 1999c,
4U\,1746$-$37; Jonker et al. 1999a, 4U\,1323$-$62; Jonker et
al. 1999b, 1E\,1724$-$3045, SLX\,1735$-$269 and GS\,1826$-$238; Barret
et al. 1999).

The QPO frequencies increase when the sources move along the tracks in
the sense of increasing inferred \mdot\ (this has been seen in more
than a dozen sources, and no counterexamples are known).  
\hide{Z sources: van der Klis et al. 1996 sco, Jonker et al. 1998 340,
Wijnands et al. 1997 17+2, Wijnands et al. 1998 5-1, Wijnands et
al. 1998 cygx2; six Z sources except GX\,349+2; references see
Table\,\ref{tab:z} and in the atoll sources M\'endez et al. 1999 1608,
van Straaten et al. 1999 0614; M\'endez \& van der Klis 1999 di Salvo
et al. 1999 1728, Bloser et al. 1999 1820, Reig et al. 1999 aql,
Boirin et al. 1999 1916, Fig.\ref{fig:1608cd}, M\'endez 1999 1636} 
On time scales of hours to a day increasing inferred \mdot\ usually
corresponds to increasing X-ray flux, so on these time scales kHz QPO
frequency usually increases with flux.  When flux systematically
decreases with inferred \mdot, as is the case in some parts of these
tracks, the frequency is expected to maintain its positive correlation
with inferred \mdot\ and hence become anticorrelated to flux, and
indeed this has been observed in the Z source GX\,17+2 (Wijnands et
al. 1997b).

So, the kHz QPOs fit well within the pre-existing Z/atoll description
of LMXB phenomenology in terms of source types and states, including
the fact that position on the tracks in X-ray color-color or
hardness-intensity diagrams (``inferred \mdot'', see \S\ref{sect:Lx})
and not X-ray flux drives the phenomenology.

\subsection{Dependence of QPO frequency on luminosity and spectrum}
\label{sect:Lx}\label{sect:parlines}

Kilohertz QPOs occur at similar frequency in sources that differ in
X-ray luminosity $L_x$ by more than 2 orders of magnitud, and the kHz
QPO frequency $\nu$ seems to be determined more by the difference
between average and instantaneous $L_x$ of a source than by $L_x$
itself (van der Klis 1997, 1998). In a plot of $\nu$ vs. \Lx\ (defined as
$4\pi d^2f_x$ with $f_x$ the X-ray flux and $d$ the distance; Ford et
al. 1999; Fig.\,\ref{fig:fordparlines}) a series of roughly parallel
lines is seen, to first order one line per source (but see below).  In
each source there is a definite relation between
\Lx\ and $\nu$, but the same relation does not apply in
another source with a different average \Lx. Instead, that source
covers the same $\nu$-range around {\it its} particular
average \Lx. This is unexplained, and must mean that in addition to
instantaneous \Lx, another parameter, related to average
\Lx, affects the QPO frequency (van der Klis 1997, 1998). Perhaps this
parameter is the neutron star magnetic field strength, which
previously, on other grounds, was hypothesized to correlate to average
\Lx\ (Hasinger \& van der Klis 1989, Psaltis \& Lamb 1998, Konar \&
Bhattacharya 1999, see also Lai 1998), but other possibilities exist
(van der Klis 1998, Ford et al. 1999).

\begin{figure}[htbp]
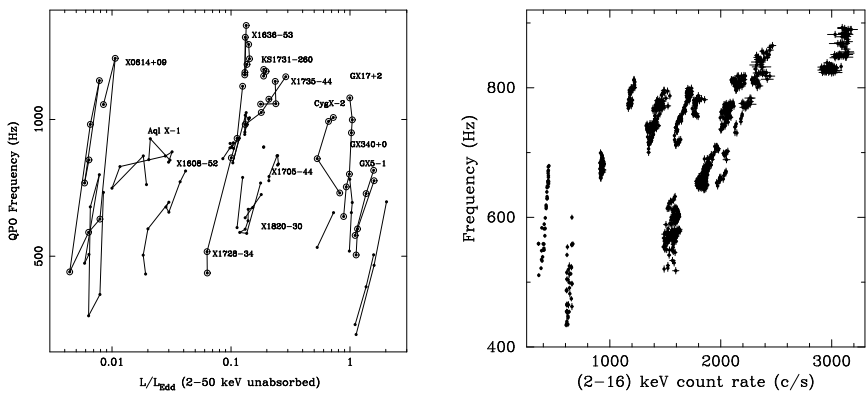

$$\psfig{figure=ford_allsources.postscript,height=2in}\qquad 
\psfig{figure=mendez_1608_apjlxxx_fig2_mess.ps,height=2in,angle=-90}$$
\caption{The parallel lines phenomenon across sources (left; Ford et
al. 1999; upper and lower kHz peaks are indicated with different
symbols) and in the source 4U\,1608$-$52 (right; M\'endez et
al. 1999, frequency plotted is
$\nu_1$). \label{fig:fordparlines}\label{fig:mendezparlines}}
\end{figure}

\begin{figure}[htbp]
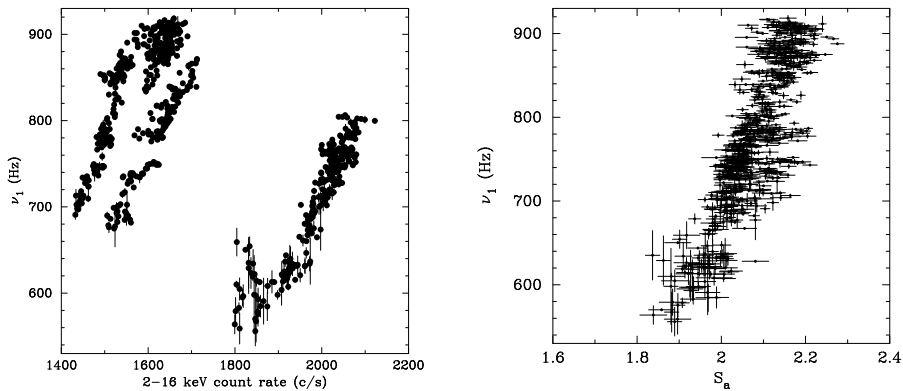

$$\psfig{figure=mendez_1728_apj517L51_fig3.ps,height=2in,angle=-90}\qquad
\psfig{figure=arev99_fig9right.ps,height=2in,angle=-90}$$
\caption{In 4U\,1728$-$34 a QPO frequency ($\nu_1$) vs. count rate
plot (left) shows no clear correlation, but instead a series of
parallel lines. When the same data are plotted vs. position in the
X-ray color-color diagram a single relation is observed. (after
M\'endez \& van der Klis 1999)
\label{fig:mendez1728}\hide{this one could go as well}
\label{fig:mendezfreqhardness}}
\end{figure}

A similar pattern of parallel lines, but on a much smaller scale,
occurs in some individual sources. When observed at different epochs,
a source produces different frequency vs. flux tracks that are
approximately parallel (GX\,5$-$1: Wijnands et al. 1998c; GX\,340+0:
Jonker et al. 1998; 4U\,0614+09: Ford et al. 1997a,b, M\'endez et
al. 1997, van Straaten et al. 1999; 4U\,1608$-$52: Yu et al. 1997,
M\'endez et al. 1998a, 1999; 4U\,1636$-$53: M\'endez 1999; 4U\,1728$-$34:
M\'endez \& van der Klis 1999; 4U\,1820$-$30: Kaaret et al. 1999b;
Aql\,X-1: Zhang et al. 1998c, Reig et al. 1999;
Figs.\,\ref{fig:mendezparlines} and \ref{fig:mendez1728}). This is
most likely another aspect of the well-known fact (e.g., van der Klis
et al. 1990, Hasinger et al. 1990, Kuulkers et al. 1994, 1996, van der
Klis 1994, 1995a) that while the properties of timing phenomena such
as QPOs are well correlated with one another and with X-ray spectral
{\it shape} as diagnosed by X-ray colors (and hence with position in
tracks in color-color diagrams) the {\it flux} correlates well to
these diagnostics only on short (hours to days) time scales and much
less well on longer time scales. This is why color-color diagrams,
independent of flux, are popular for parametrizing spectral
variability in these sources. Similarly to other timing parameters,
kHz QPO frequency correlates much better with position on the track in
the color-color diagram (Fig.\,\ref{fig:mendezfreqhardness}) than with
flux. Correlations of frequency with parameters describing spectral
shape such as blackbody flux (Ford et al. 1997b) or power-law slope
(Kaaret et al. 1998) in a two-component spectral model are also much
better than with flux.

Usually, all this has been interpreted by saying that apparently
inferred accretion rate (\S\ref{sect:zatol}) governs both the timing
and the spectral properties, but not flux (e.g., van der Klis
1995a). Of course, energy conservation suggests total \mdot\ and flux
should be well correlated. Perhaps, inferred \mdot\ is not the total
\mdot\ but only one component of it, i.e., that through the disk,
while there is also a radial inflow (e.g. Fortner et al. 1989, Kuulkers
\& van der Klis 1995, Wijnands et al. 1996b, Kaaret et al. 1998), maybe
there are large and variable anisotropies or bolometric corrections in
the emission, so that the flux we measure is not representative for the
true luminosity (e.g., van der Klis 1995a), or possibly mass outflows
destroy the expected correlation by providing sinks of both mass and
(kinetic) energy (e.g., Ford et al. 1999). The true explanation is
unknown. We do not even know if the two different parallel-lines
phenomena (across sources and within individual sources;
Fig.\,\ref{fig:fordparlines}) have the same origin. It is possible that
the quantity that everything depends on is {\it not} \mdot, but some
other parameter such as inner disk radius $r_{in}$.

\subsection{Peak separation and burst oscillation frequencies} 
\label{sect:deltanu}

\begin{figure}[htbp]
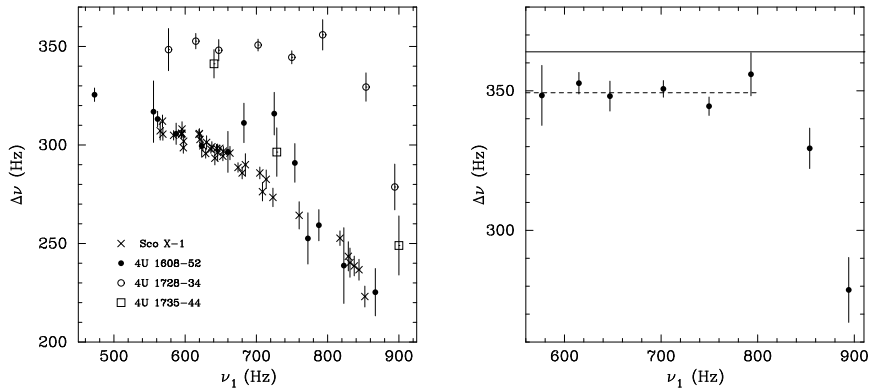

$$\psfig{figure=mendez_4srces_deltanu.ps,height=2in,angle=-90}\qquad 
\psfig{figure=mendez_1728_deltanu.ps,height=2in,angle=-90}
$$
\caption{Left: the variations in kHz QPO peak separation as a function 
of the lower kHz frequency. Right: the data on 4U\,1728$-$34. The burst
oscillation frequency is indicated by a horizontal drawn line (M\'endez
\& van der Klis 1999).
\label{fig:deltanuchanges}}
\end{figure}

In interpretations (\S\ref{sect:formulas}, \ref{sect:spm}) where the
kHz peak separation $\nu_2-\nu_1=\Delta\nu$ is the neutron star spin
frequency $\nu_s$ one expects $\Delta\nu$ to be approximately
constant. This is not the case (Fig.\,\ref{fig:deltanuchanges}). In
Sco\,X-1 (van der Klis 1996c, 1997b), 4U\,1608$-$52 (M\'endez et
al. 1998a,b), 4U\,1735$-$44 (Ford et al. 1998b), 4U\,1728$-$34 (M\'endez
\& van der Klis 1999) and marginally also 4U\,1702$-$43 (Markwardt et
al. 1999a) the separation $\Delta\nu$ decreases considerably when the
kHz QPO frequencies increase. Other sources may show similar
$\Delta\nu$ variations (see also Psaltis et al. 1998). In a given
source the $\nu_1$, $\nu_2$ relation seems to be reproducible.

That $\nu_{burst}$, by interpretation close to $\nu_s$
(\S\ref{sect:burstosc}), is close to $\Delta\nu$ (or $2\Delta\nu$) is
the most direct evidence 
for a beat frequency interpretation of the kHz QPOs (\S\ref{sect:formulas}). The evidence for
this is summarized in Table~\ref{tab:beatevidence}, where the highest
asymptotic burst frequency (likely to be closest to the spin frequency;
\S\ref{sect:burstosc}) is compared to the largest well-measured
$\Delta\nu$ in the five sources where both have been measured. The
frequency ratio may cluster at 1 and 2, but a few more examples are
clearly needed; in two sources discrepancies of \about15\% occur
between $\nu_{burst}$ and $2\Delta\nu$. Of these five sources,
4U\,1728$-$34 also has a measured $\Delta\nu$ variation
(Fig.\,\ref{fig:deltanuchanges}).  When the QPO frequency drops,
$\Delta\nu$ increases, to saturate 4\% below the burst oscillation
frequency (M\'endez \& van der Klis 1999). It certainly seems here as
if the kHz QPO separation ``knows'' the value of the burst oscillation
frequency.
%
%
\begin{table}[htbp]
\caption{Commensurability of kHz QPO and burst oscillation frequencies
\label{tab:beatevidence}} 
\footnotesize
\begin{center}
\begin{tabular}{|lccccc|}
\hline 
Source & Highest       & Highest     & Ratio                     & Discrepancy & References\\
       & $\nu_{burst}$ & $\Delta\nu$ & ($\nu_{burst}/\Delta\nu$) &             &           \\
       &          (Hz) & (Hz)        &                           & (\%)        &           \\
\hline
4U\,1702$-$43 & 330.55\p0.02  & 344\p7      & 0.96\p0.02                & $-$4\p2 & 1\\
4U\,1728$-$34 & 364.23\p0.05  & 349.3\p1.7  & 1.043\p0.005      & +4.3\p0.5 & 2\\
\hline
KS\,1731$-$260& 523.92\p0.05  & 260\p10     & 2.015\p0.077    & +0.7\p3.8 &3 \\
Aql\,X-1      & 548.9         & 241\p9$^a$  & 2.28\p0.09$^a$          & +14\p5$^a$ & 4\\
4U\,1636$-$53 & 581.75\p0.13  & 254\p5      & 2.29\p0.04                & +15\p2  & 5\\
\hline
\end{tabular}
\end{center}
\vbox{$^a$ Based on a marginal detection (see Table\,\ref{tab:atoll}). References: [1] Markwardt et
al. 1999a [2] M\'endez \& van der Klis 1999 [3] Wijnands \& van der Klis
1997 [4] M. M\'endez et al. 1999 in prep. [5] M\'endez et al. 1998c}
\end{table}
\subsection{Strong gravity and dense matter}\label{sect:relativity}

Potentially kHz QPOs can be used to constrain neutron-star masses and
radii, and to test general relativity. Detection of the predcited
innermost stable circular orbit (ISCO, \S\ref{sect:formulas}) would
constitute the first direct detection of a strong-field
general-relativistic effect and prove that the neutron star is smaller
than the ISCO. This possibility has fascinated since the beginning and
was discussed well before the kHz QPOs were found (Klu\'zniak and
Wagoner 1985, Paczynski 1987, Klu\'zniak et al. 1990, Klu\'zniak \&
Wilson 1991, Biehle \& Blandford 1993).  If a kHz QPO is due to
orbital motion around the neutron star, its frequency can not be
larger than the frequency at the ISCO. Kaaret et al. (1997) proposed
that kHz QPO frequency variations that then seemed uncorrelated to
X-ray flux in 4U\,1608$-$52 and 4U\,1636$-$53 were due to orbital
motion near the ISCO and from this derived neutron star masses of
\about2\msun. However, from more detailed studies (\S\ref{sect:Lx}) it
is clear now that on short time scales frequency does in fact
correlate to flux in these sources.

The maximum kHz QPO frequencies observed in each source are
constrained to a narrow range. The 12 atoll sources with twin peaks
have maximum $\nu_2$ values in the range 1074--1329\,Hz; among the Z
sources there are two cases of much lower maximum $\nu_2$ values
($<$900\,Hz in GX\,5$-$1 and GX\,340+0), while the other four fit
in. Zhang et al. (1997b) proposed that this narrow distribution is
caused by the limit set by the ISCO frequency, which led them to
neutron star masses near 2\msun\ as well. It is in principle possible
that the maximum is set by some other limit on orbital radius (e.g.,
the neutron star surface), or that it {\it is} caused by the ISCO, but
the frequency we observe is not orbital, in which cases no mass
estimate can be made.

\begin{figure}[htbp]
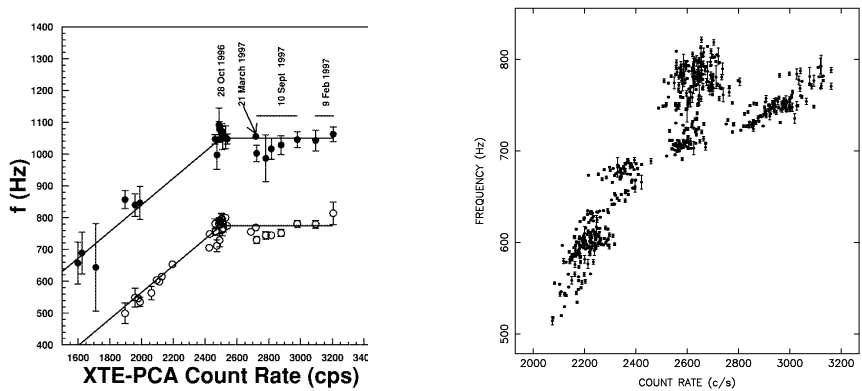

$$\psfig{figure=1820_zhang_apjl500_fig4_saturation.postscript,height=2.2in,angle=0}\qquad
\psfig{figure=1820_mendez_freqcntrate_inprep.postscript,height=2.in,angle=0}$$
\caption{Left: evidence for a leveling off of the kHz QPO frequency with
count rate in 4U\,1820$-$30. (Zhang et al. 1998b) Right: a larger data
set (see also Kaaret et al. 1999b) of 4U\,1820$-$30 at higher
resolution showing only the lower peak's frequency $\nu_1$. (M. M\'endez et
al. in prep.). \label{fig:1820ISCO}}
\end{figure}

Miller et al. (1996, 1998a) suggested that when the inner edge of the
accretion disk reaches the ISCO, the QPO frequency might level off and
remain constant while \mdot\ continues rising. Later, an apparent
leveling off at $\nu_2$=1060$\pm$20\,Hz with X-ray count rate was
found in 4U\,1820$-$30 (Zhang et al. 1998b; Fig.~\ref{fig:1820ISCO},
left).  If this is the orbital frequency at the ISCO, then the neutron
star has a mass mass of \about2.2\msun\ and is smaller than its ISCO,
and many equations of state are rejected. The leveling off is also
observed as a function of X-ray flux and color (Kaaret et al. 1999b)
and position along the atoll track (Bloser et al. 1999). However, the
frequency vs. flux relations are known in other sources
(\S\ref{sect:Lx}) to be variable, and in 4U\,1820$-$30 the leveling
off seems not to be reproduced in the same way in all data sets
(M. M\'endez et al. in prep.; Fig.\,\ref{fig:1820ISCO}, right). It may also
be more gradual in nature than originally suggested. No evidence for a
similar saturation in frequency was seen in other sources, and most
reach higher frequencies. Possibly, the unique aspects of the
4U\,1820$-$30 system are related to this (it is an 11-min binary in a
globular cluster with probably a pure He companion star; Stella et
al. 1987a).

\begin{figure}[htbp]
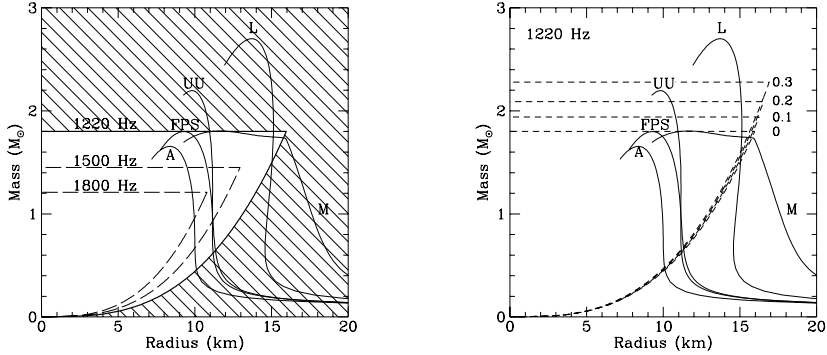
\vskip-5mm
$$\psfig{figure=millerlambpsaltis_fig14.postscript,height=2.25in,angle=0}
\psfig{figure=millerlambpsaltis_fig15.postscript,height=2.25in,angle=0}$$
\caption{Constraints on the mass and radius of neutron stars from
the detection of orbital motion with the frequencies indicated.
Graphs are for negligible neutron star angular momentum (left) and for
the values of $j=cJ/GM^2$ (\S\ref{sect:khzqpos}) indicated (right).
Mass-radius relations for some representative EOSs are shown. (Miller
et al. 1998a) \label{fig:wedge}}
\end{figure}

If a kHz QPO peak at frequency $\nu$ corresponds to stable Keplerian
motion around a neutron star, one can immediately set limits on
neutron star mass $M$ and radius $R$ (Miller et al. 1998a). For a
Schwarzschild geometry: (1) the radius $R$ of the star must be smaller
than the radius $r_K$ of the Keplerian orbit: $R < r_K =
(GM/4\pi^2\nu^2)^{1/3}$, and (2) the radius of the ISCO
(\S\ref{sect:formulas}) must {\it also} be smaller than $r_K$, as no
stable orbit is possible within this radius: $r_{ISCO} = 6GM/c^2 <
(GM/4\pi^2\nu^2)^{1/3}$ or $M < c^3/(2\pi6^{3/2}G\nu)$.  Condition (1)
is a mass-dependent upper limit on $R$, and condition (2) an upper
limit on $M$; neither limit requires detection of orbital motion at
the ISCO.  Fig.\,\ref{fig:wedge} (left) shows these limits in the
neutron star mass-radius diagram for $\nu=1220$\,Hz, plus an
indication of how the excluded area ({\it hatched}) shrinks for higher
values of $\nu$. The currently highest value of $\nu_2$, identified in
most models with the orbital frequency, is 1329\p4\,Hz (van Straaten
et al. 1999), so the hardest equations of state are beginning to be
imperiled by the method.  Corrections for frame dragging (shown to
first order in $j$ [\S\ref{sect:formulas}, right] in
Fig.\,\ref{fig:wedge}) expand the allowed region. They depend on the
neutron star spin rate $\nu_s$, and somewhat on the neutron star
model, which sets the relation between $\nu_s$ and angular momentum.
For 1329\,Hz the above equations imply $M<1.65$\msun\ and
$R_{NS}<12.4$\,km; with corrections for a 300\,Hz spin these numbers
become 1.9\msun\ and 15.2\,km (van Straaten et al. 1999). Calculations
exploring to what extent kHz QPOs constrain the EOS have further been
performed by Miller et al. (1998b), Datta et al. (1998), Akmal et
al. (1998), Klu\'zniak (1998), Bulik et al. (1999), Thampan et
al. (1999), Li et al. (1999b), Schaab \& Weigel (1999) and Heiselberg
\& Hjorth-Jensen (1999).

If $\Delta\nu$ is near $\nu_s$ then the 18 neutron stars where this
quantity has been measured all spin at frequencies between \about240 and
\about 360\,Hz, a surprisingly narrow range. If the stars spin at
the magnetospheric equilibrium spin rates (e.g., Frank et al. 1992)
corresponding to their current luminosities \Lx, this would imply an
unlikely, tight correlation between \Lx\ and neutron-star
magnetic-field strength (White \& Zhang 1997; note that a similar
possibility came up in the discussion about the uniformity of the QPO
frequencies themselves, \S\ref{sect:Lx}).  White \& Zhang (1997)
propose that when $r_M$ is small, as is the case here, it depends only
weakly on accretion rate (as it does in some inner disk models;
cf. Ghosh \& Lamb 1992, Psaltis \& Chakrabarty 1999). Another
possibility is that the spin frequency of accreting neutron stars is
limited by gravitational radiation losses (Bildsten 1998a, Andersson
et al. 1999, see also Levin 1999). If so, then gravitational radiation
is transporting angular momentum out as fast as accretion is
transporting it in, making these sources the brightest gravitational
radiation sources in the sky. They would produce a periodic signature
at the neutron star spin frequency, which would facilitate their
detection.

\subsection{Other kilohertz QPO properties}\label{sect:otherprops}

\begin{figure}[htbp]
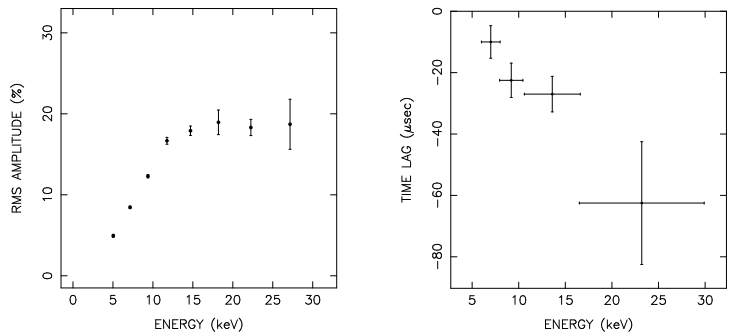

\begin{center}
\begin{tabular}{c}
\psfig{figure=arev99_fig13_left.postscript,height=1.7in} \qquad
\psfig{figure=apjl_xte_vaughan_khzqpolags.postscript,height=1.7in}
\end{tabular}
\caption{Energy dependence (left) and time lags (right) of
the lower kHz peak in 4U\,1608$-$52. (Berger et al. 1996, Vaughan et
al. 1997, 1998) \label{fig:energydependence}}
\end{center}
\end{figure}

The amplitudes of kHz QPOs increase strongly with photon energy
(Fig.\,\ref{fig:energydependence}). In similar X-ray photometric bands
the QPOs tend to be weaker in the more luminous sources, with
2--60\,keV amplitudes ranging from as high as 15\% (rms) in
4U\,0614+09, to typically a few \% (rms) at their strongest in the Z
sources. At high energy amplitudes are much higher (e.g., 40\% rms
above 16\,keV in 4U\,0614+09; M\'endez et al. 1997).  Fractional rms
usually decreases with inferred \mdot\ (e.g., van der Klis et
al. 1996b, 1997b, Berger et al. 1996) but more complex behavior is
sometimes seen (e.g., di Salvo et al. 1999). The measured widths of
QPO peaks are affected by variations in centroid frequency during the
measurement, but typical values in the Z sources are 50--200\,Hz. In
the atoll sources, the upper peak usually has a width similar to this,
although occasionally peaks as narrow as 10 or 20\,Hz have been
measured (eg. Wijnands \& van der Klis 1997, Wijnands et al. 1998a,b)
but the lower peak is clearly much narrower. It is rarely as wide as
100\,Hz (M\'endez et al. 1997) and sometimes as narrow as 5\,Hz (e.g.,
Berger et al. 1996, Wijnands et al. 1998b). Various different peak
width vs. inferred \mdot\ relations have been seen, but there seems to
be a tendency for the upper peak to become narrower as its frequency
increases.

A strong model constraint is provided by the time lags between kHz QPO
signals in different energy bands (\S\ref{sect:fft}). Time-lag
measurements require very high signal-to-noise ratios (Vaughan et al.
1997), and have mostly been made in the very significant lower peaks
observed in some atoll sources. Finite lags of 10--60$\mu$sec occur in
these peaks (Vaugan et al. 1997, 1998, Kaaret et al. 1999a, Markwardt et
al. 1999b; see Lee \& Miller 1998 for a calculation of Comptonization
lags relevant to kHz QPOs). Contrary to the initial report, the
low-energy photons lag the high-energy ones (these are ``soft lags'') by
increasing amounts as the photon energy increases
(Fig.\,\ref{fig:energydependence}). The lags are of opposite sign to
those expected from the inverse Compton scattering thought to produce
the hard spectral tails of these sources (e.g., Barret \& Vedrenne
1994), and correspond to light travel distances of only 3--20\,km. From
this it seems more likely that the lags originate in the QPO production
mechanism than in propagation delays. Markwardt et al. (1999b) reported
a possible hard lag in an atoll-source upper peak.

\section{Correlations with low-frequency timing phenomena and with
other sources}
\label{sect:vuiltjes}

\subsection{Black hole candidates}\label{sect:bhc}

The only oscillations with frequencies exceeding 10$^{2.5}$\,Hz
(\S\ref{sect:intro}) known in black hole candidates (BHCs) are,
marginally, the 100--300\,Hz oscillations in GRO\,J1655$-$40, and
XTE\,J1550$-$564, and those reported very recently in Cyg\,X-1 and
4U\,1630$-$47. An oscillation near 67\,Hz observed in GRS\,1915+105 is
usually discussed together with these QPOs, although it is not clear
that it is related.  Usually, the phenomenology accompanying these
high-frequency QPOs (spectral variations, lower-frequency variations) is
complex.

The 67\,Hz QPO in GRS\,1915+105 (Morgan et al. 1997, Remillard \&
Morgan 1998) varies by only a few percent in frequency when the X-ray
flux varies by a factor of several. It is relatively coherent, with Q
usually around 20 (but sometimes dropping to 6; Remillard et
al. 1999a), has an rms amplitude of about 1\%, and the signal at high
photon energy lags that at lower energy by up to 2.3 radians (Cui
1999). The 300\,Hz QPO in GRO\,J1655$-$40 (Remillard et al. 1999b) was
seen only when the X-ray spectrum was dominated by a hard power law
component. This feature was relatively broad (Q\about4) and weak
(0.8\% rms), and did not vary in frequency by more than
\about30\,Hz. The 185--285\,Hz QPO in XTE\,J1550$-$564 (Remillard et
al. 1999c) shows considerable variations in frequency (Homan et
al. 1999b, Remillard et al. 1999c). It is seen in similar X-ray
spectral conditions as the 300\,Hz QPO in GRO\,J1655$-$40, has
similarly low amplitudes and 3$<$Q$<$10. Recent reports indicate that
QPOs in this frequency range also occur in Cyg\,X-1 and 4U\,1630$-$47
(Remillard 1999a,b).

The fact that the 300 and 67\,Hz oscillations were constant in
frequency (different from anything then known to occur in neutron
stars) triggered interpretations where these frequencies depend mostly
on black-hole mass and angular momentum and only weakly on luminosity,
such as orbital motion at the ISCO (Morgan et al. 1997),
Lense-Thirring precession there (Cui et al. 1998c, see also Merloni et
al. 1999) or trapped-mode disk oscillations (Nowak et al. 1997).
However, the variations in frequency of the QPO in XTE\,J1550$-$564
have cast some doubt on the applicability of such models.

As is well known, strong similarities exist with respect to many
spectral and timing phenomena between low-magnetic-field neutron stars
and BHCs (e.g., van der Klis 1994a,b; \S\ref{sect:slowqpo}). While the
100--300\,Hz oscillations may be related to the kHz QPOs observed in
neutron stars (Psaltis et al. 1999a; see \S\ref{sect:slowqpo}), there
could also be a relation with the recently reported relatively stable
QPO peaks near 100\,Hz in 4U\,0614+09 and 4U\,1728$-$34 (van Straaten
et al. 1999, di Salvo et al. 1999), which are clearly distinct from
kHz QPOs. More work clearing up the exact phenomenology and more
observations of black-hole transients leading to more examples of high
frequency QPOs are clearly needed.

\subsection{Cen X-3}\label{sect:cenx3}

The detection of QPO features near 330 and 760\,Hz in the 4.8\,s
accreting pulsar Cen\,X-3 was recently reported by Jernigan et
al. (1999). This is the first report of millisecond oscillations from
a high-magnetic-field (\about10$^{12}$\,Gauss) neutron star. The QPO
features are quite weak. Jernigan et al. (1999) carefully discuss the
instrumental effects, which are a concern at these low power levels
and interpret their results in terms of the photon bubble model
(\S\ref{sect:pbm}).

\subsection{Low-frequency phenomena}\label{sect:slowqpo}

\begin{figure}[htbp]
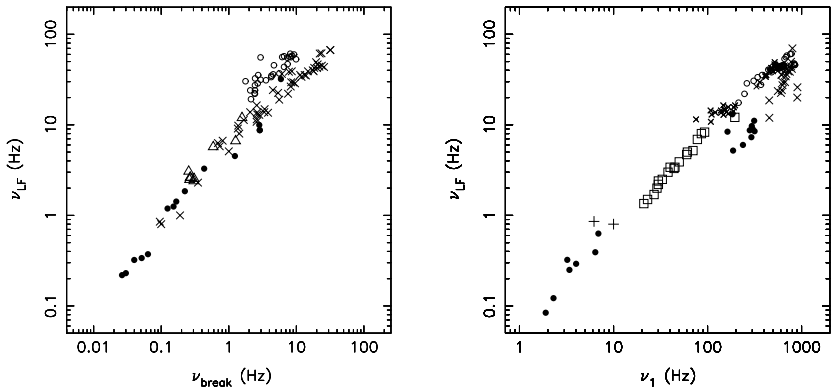

$$\psfig{figure=wijands_vdklis_brqpo.postscript,height=2.in}\qquad
\psfig{figure=psaltis_belloni_vdklis_f1flf.postscript,height=2.in}
$$
\caption{The frequencies of various QPO and broad noise
components seen in accreting neutron stars and black holes plotted
vs. each other suggest that components similar to Z-source HBO
($\nu_{LF}$) and the lower kHz peak ($\nu_1$) occur in all these
sources and span a very wide range in frequency. Left: $\nu_{LF}$
vs. noise break frequency (after Wijnands \& van der Klis 1999a);
right: $\nu_{LF}$ vs. $\nu_1$ (after Psaltis et al. 1999a). Filled
circles represent black hole candidates, open circles Z sources,
crosses atoll sources (the smaller crosses in the right hand frame are
data of 4U\,1728$-$34 where $\nu_1$ was obtained from
$\nu_1=\nu_2-363$\,Hz), triangles the millisecond pulsar
SAX\,J1808.4$-$3658, pluses faint burst sources and squares (from
Shirey et al. 1996) Cir\,X-1. \label{fig:psaltisbellonivdklis}
\label{fig:wijnandsbbn}}
\end{figure}

Low-frequency ($<$100\,Hz) QPOs have been studied in accreting neutron
stars and black hole candidates since the 1980's, mostly with the
EXOSAT and Ginga satellites (see van der Klis 1995a). Two different
ones were known in the Z sources, the 6-20\,Hz so-called normal and
flaring-branch oscillation (NBO; Middleditch \& Priedhorsky 1986) and
the 15--60\,Hz so-called horizontal branch oscillation (HBO; van der
Klis et al. 1985). The frequency of the HBO turned out to correlate
well to those of the kHz QPOs (Fig.\,\ref{fig:psaltisbellonivdklis};
van der Klis et al. 1997b, Wijnands et al. 1997b,1998a,c, Jonker et
al. 1998, 1999c), and the same is true for the NBO in Sco\,X-1 (van der
Klis et al. 1996b).  Broad power-spectral bumps and, rarely,
low-frequency QPOs were know in atoll sources as well (Lewin et
al. 1987, Stella et al. 1987b, Hasinger \& van der Klis 1989, Dotani
et al. 1989, Makishima et al. 1989, Yoshida et al. 1993). With RXTE,
QPOs similar to HBO are often seen (Strohmayer et al. 1996b, Wijnands
\& van der Klis 1997, Wijnands et al. 1998b, Homan et al. 1998a) and
their frequencies also correlate well to kHz QPO frequency
(Fig.\,\ref{fig:psaltisbellonivdklis}; Stella \& Vietri 1998, Ford
\& van der Klis 1998, Markwardt et al. 1999a, van Straaten et
al. 1999, di Salvo et al. 1999, Boirin et al. 1999).  It is not sure
yet wether these atoll QPOs are physically the same as HBO in Z
sources but this seems likely. 

\begin{figure}[htbp]
$$\psfig{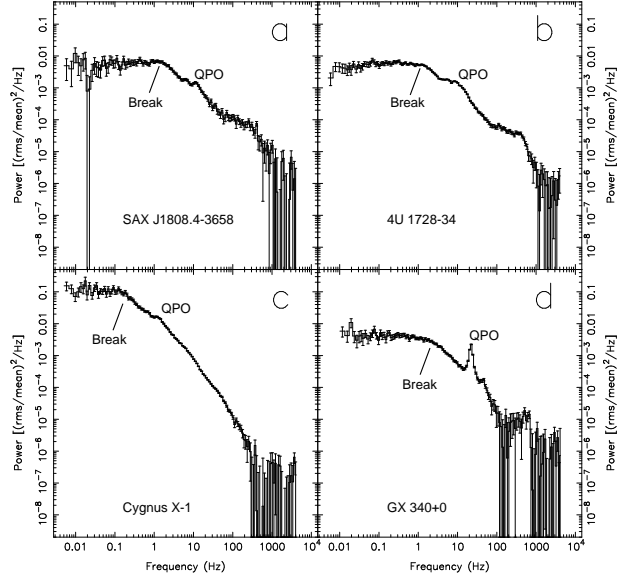}$$
\caption{Broad-band power spectra of, respectively, the millisecond 
pulsar, an atoll source, a black-hole candidate and a Z
source. (Wijnands \& van der Klis 1999a) \label{fig:wijnandsspectra}}
\end{figure}

It is possible that these correlations arise just because all QPO
phenomena depend on a common parameter (e.g. inferred \mdot,
\S\ref{sect:Lx}), but Stella \& Vietri (1998) proposed that their
origin is a physical dependence of the frequencies on one another
(\S\ref{sect:rpm}). In their relativistic precession model the HBO and
the similar-frequency QPOs in the atoll sources {\it are} the same
phenomenon, and their frequency $\nu_{LF}$ is predicted to be
proportional to ${\nu_2}^2$. This is indeed approximately true in all
Z sources except Sco\,X-1 (Psaltis et al. 1999b), as well as in the
atoll sources (references above). Psaltis et al. (1999b) argue that a
combination of the sonic-point and magnetospheric beat-frequency
models can explain these correlations as well (\S\ref{sect:spm}).

Additional intriguing correlations exist between kHz QPOs and
low-frequency phenomena which may link neutron stars and BHCs.  It is
useful to first examine a correlation between two low-frequency
phenomena. At low \mdot, BHCs and atoll sources (van der Klis 1994a and
references therein), the millisecond pulsar SAX\,J1808.4-3658
(\S\,\ref{sect:mspulses}; Wijnands \& van der Klis 1998c),
and perhaps even Z sources have very similar power spectra
(Fig.~\ref{fig:wijnandsspectra}; Wijnands \& van der Klis 1999a),
with a broad noise component that shows a break at low frequency and
often a QPO-like feature above the break. Break and QPO frequency both
vary in excellent correlation (Fig.~\ref{fig:wijnandsbbn}, left), and
similarly in neutron stars and BHCs. This suggests that (with the
possible exception of the Z sources, which are slightly off the main
relation) these two phenomena are the same in neutron stars and black
holes. This would exclude spin-orbit beat-frequency models and any
other models requiring a material surface, an event horizon, a
magnetic field, or their absence, and would essentially imply the
phenomena are generated in the accretion disk around {\it any}
low-magnetic field compact object.

The good correlations between kHz QPOs and low-frequency phenomena in
Z and atoll sources suggest that kHz QPOs might also fit in with
schemes linking neutron stars and BHCs (\S\ref{sect:bhc}). However,
no twin kHz QPOs have been reported from BHCs.  Psaltis et al. (1999a)
pointed out that many Z and atoll sources, the peculiar source
Cir\,X-1 and a few low luminosity neutron stars and BHCs sometimes
show two QPO or broad noise phenomena whose centroid frequencies, when
plotted vs. each other, seem to line up
(Fig.~\ref{fig:psaltisbellonivdklis}, right). This suggests that
perhaps ``kHz QPOs'' {\it do} occur in BHCs, but as features at
frequencies below 50\,Hz. These features have very low Q, and although
the data are suggestive they are not conclusive. The implication would
be that the lower kHz QPO peak (whose frequency is the one that lines
up with those seen in the BHCs) is not unique to neutron stars, but a
feature of disk accretion not related to neutron star spin. The
coincidence of kHz QPO frequencies with burst oscillation frequencies
(\S\ref{sect:burstosc}, \S\ref{sect:formulas}) would then require some
other explanation. Orbital motion in the disk would remain an
attractive interpretation for some of the observed frequencies. Stella
et al. (1999b) showed that for particular choices of neutron star and
black hole angular momenta their relativistic precession model can fit
these data.  The phenomenology is quite complex; in particular, no way
has been found yet to combine the Wijnands \& van der Klis (1999a)
work with the Psaltis et al. (1999a) results in a way that works
across all source types. New low-frequency phenomena are still being
discovered as well (di Salvo et al. 1999, Jonker et al. 1999c).

\section{Kilohertz QPO models}\label{sect:models}

The possibility to derive conclusions of a fundamental nature has led
to a relatively large number of models for kHz QPOs. Most, but not all
of these involve orbital motion around the neutron star. It is beyond
the scope of the present work to provide an in-depth discussion of
each model. Instead, I point out some of the main issues, and 
provide pointers to the literature.

In early works, the magnetospheric beat-frequency model was implied
when beat-frequency models were mentioned (e.g., van der Klis et
al. 1996b, Strohmayer et al. 1996c), but this model has not recently
been applied much to kHz QPOs; see however Cui et al. (1998a). Most
prominent recently have been the sonic point beat-frequency model of
Miller et al. (1996, 1998a) and the relativistic precession model of
Stella and Vietri (1998, 1999), but also the the photon bubble model
(Klein et al. 1996b) and the disk transition layer models (Titarchuk et
al. 1998, 1999) have been strongly argued for. Additional disk models
have been proposed as well (\S\ref{sect:disk}). Neutron star
oscillations have been considered (Strohmayer et al. 1996c, Bildsten
et al. 1998, Bildsten \& Cumming 1998), but probably can not produce
the required combination of high frequencies and rapid changes in
frequency.

Most models have evolved in response to new observational results. The
sonic point model was modified to accommodate the observed deviations
from a pure beat-frequency model (\S\ref{sect:deltanu}; Lamb \&
Miller 1999, Miller 1999c), the relativistic precession model
initially explained the lower kHz peak as a spin/orbit beat frequency
(Stella \& Vietri 1998) and only later by apsidal motion (Stella \&
Vietri 1999), the photon bubble model is based on numerical
simulations which in time have become more similar to what is observed
(R Klein, priv. comm.), and also the disk transition layer models have
experienced considerable evolution (e.g., Osherovich \& Titarchuk
1999).

The relativistic precession model makes the strongest predictions with
respect to observable quantities and hence allows the most direct
tests, but the near-commensurability of kHz QPO and burst oscillation
frequencies (\S\ref{sect:deltanu}) is unexplained in that model. The
sonic-point model provides specific mechanisms to modulate the X-rays
and to make the frequency vary with mass-accretion rate. Other models
usually discuss at least one of these issues only generically (usually
in terms of self-luminous and/or obscuring blobs, and arbitrary
preferred radii in the accretion disk).

\subsection{The sonic point beat-frequency model}\label{sect:spm}

Beat-frequency models involve orbital motion at some preferred radius
in the disk (\S\ref{sect:formulas}).  A beat-frequency model which
uses the magnetospheric radius $r_M$ was proposed by Alpar \& Shaham
(1985) to explain the HBO in Z sources (\S\ref{sect:slowqpo}; see also
Lamb et al. 1985). In the Z sources HBO and kHz QPOs have been seen
{\it simultaneously}, so at least one additional model is required.

Miller et al. (1996, 1998a) suggest to continue to use the
magnetospheric model for the HBO (see also Psaltis et al. 1999b), and
for the kHz QPOs propose the sonic-point beat-frequency
model. In this model the preferred radius is the sonic radius
$r_{sonic}$, where the radial inflow velocity becomes supersonic. This
radius tends to be near $r_{ISCO}$ (\S\ref{sect:formulas}) but
radiative stresses change its location, as required by the observation
that the kHz QPO frequencies vary.  Comparing the HBO and kHz QPO
frequencies, clearly $r_{sonic}\ll r_M$, so part of the accreting
matter must remain in near-Keplerian orbits well within $r_M$.
\begin{figure}[htbp] \vskip-5mm
\begin{center}
\begin{tabular}{c}
\psfig{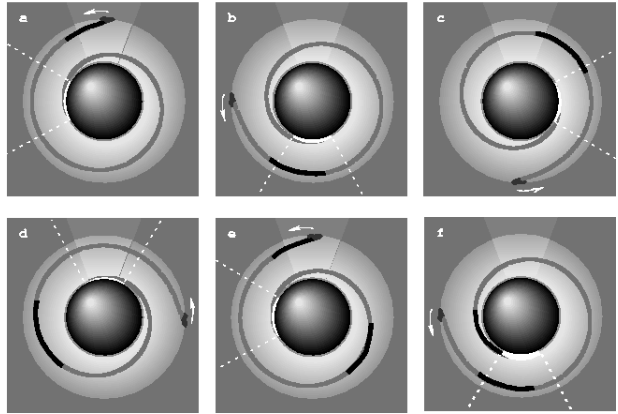}
\end{tabular}
\caption{The clump with its spiral flow, the emission from the flow's 
footpoint (dashed lines) and the clump's interaction with the pulsar
beam (lighter shading) in the Miller et al. (1998a)
model. \label{fig:spiralflow}}
\end{center}
\end{figure}

At $r_{sonic}$ orbiting clumps form whose matter gradually accretes
onto the neutron star following a fixed spiral-shaped trajectory in
the frame corotating with their orbital motion
(Fig.\,\ref{fig:spiralflow}). At the ``footpoint''of a clump's spiral
flow the matter hits the surface and emission is enhanced. The
footpoint travels around the neutron star at the clump's orbital
angular velocity, so the observer sees a hot spot move around the
surface with the Keplerian frequency at $r_{sonic}$. This produces the
upper kHz peak at $\nu_2$. The high Q of the QPO implies that all
clumps are near one precise radius and live for several 0.01 to
0.1\,s, and allows for relatively little fluctuations in the spiral
flow. The beat frequency at $\nu_1$ occurs because a beam of X-rays
generated by accretion onto the magnetic poles sweeps around at the
neutron star spin frequency $\nu_s$ and hence irradiates the clumps at
$r_{sonic}$ once per beat period, which modulates, at the beat
frequency, the rate at which the clumps provide matter to their spiral
flows and consequently the emission from the footpoints.

So, the model predicts $\Delta\nu = \nu_2-\nu_1$ to be constant at
$\nu_s$, contrary to observations (\S\ref{sect:deltanu}). However, if
the clumps' orbits gradually spiral down, then the observed beat
frequency will be higher than the actual beat frequency at which beam
and clumps interact, because then during the clumps' lifetime the
travel time of matter from clump to surface gradually diminishes. This
puts the lower kHz peak closer to the upper one, and thus decrease
$\Delta\nu$, more so when at higher $L_x$ due to stronger radiation
drag the spiralling-down is faster, as observed (Lamb \& Miller 1999,
Miller 1999). As the exact way in which this affects the
relation between the frequencies is hard to predict, this makes
testing the model more difficult. A remaining test is that a number of
specific {\it additional} frequencies is predicted to arise from the
beat-frequency interaction (Miller et al. 1998a), which are different
from additional frequencies in, e.g., the relativistic precession
model.

\subsection{The relativistic precession model}\label{sect:rpm}

Inclined eccentric free-particle orbits around a spinning neutron star
show both nodal precession (a wobble of the orbital plane) due to
relativistic frame dragging (Lense \& Thirring 1918), and
relativistic periastron precession similar to Mercury's (Einstein
1915). The relativistic precession model (Stella \& Vietri 1998,
1999) identifies $\nu_2$ with the frequency of an orbit in the disk
and $\nu_1$ and the frequency $\nu_{LF}$ of one of the observed
low-frequency (10--100\,Hz) QPO peaks (\S\ref{sect:slowqpo}) with,
respectively, periastron precession and nodal precession of this
orbit.

\begin{figure}[htbp]
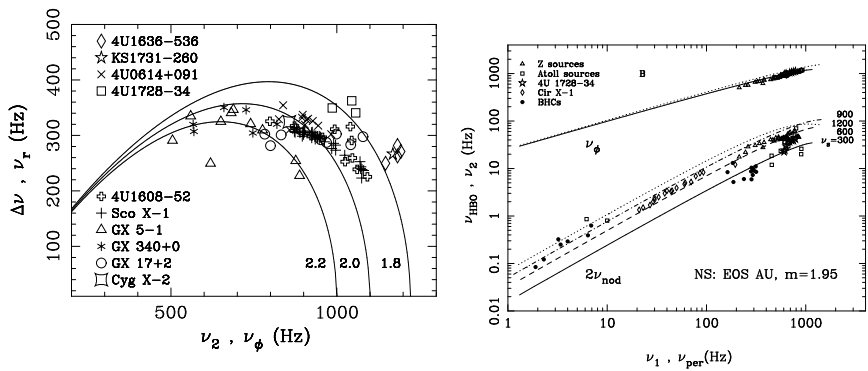

$$\psfig{figure=stella_vietri_fig1.postscript,height=1.9in,angle=-90}
\psfig{figure=stella_vietri_morsinkfig.postscript,height=1.7in,angle=-90}$$
\caption{Predicted relations between $\nu_2$ and $\Delta\nu$ (left)
and $\nu_1$ and $\nu_{LF}$ as well as $\nu_2$ (right) in the
relativistic precession model compared with observed values. (Stella
\& Vietri 1999, Stella et al. 1999b). See
\S\ref{sect:slowqpo} for a discussion of the data in the right-hand
frame. \label{fig:rpm}}
\end{figure}

To lowest order, the relativistic nodal precession is
$\nu_{nod}=8\pi^2I\nu_2^2\nu_s/c^2M$ and the relativistic periastron
precession causes the kHz peak separation to vary as
$\Delta\nu=\nu_2(1-6GM/rc^2)^{1/2}$, where $I$ is the star's moment of
inertia and $r$ the orbital radius (Stella et al. 1998, 1999; see
also Markovi\'c \& Lamb 1998).  Stellar oblateness affects both
precession rates and must be corrected for (Morsink \& Stella 1999,
Stella et al. 1999b).  For acceptable neutron star parameters, there
is an approximate match (e.g., Fig.\,\ref{fig:rpm}) with the observed
$\nu_1$, $\nu_2$ and $\nu_{LF}$ relations if $\nu_{LF}$ is {\it twice}
(or perhaps sometimes four times) the nodal precession frequency,
which could in principle arise from a warped disk geometry (Morsink
\& Stella 1999).  In this model the neutron star spin frequencies do
not cluster in the 250--350\,Hz range (\S\ref{sect:relativity}) and
$\Delta\nu$ and $\nu_{burst}$ are not expected to be equal as in
beat-frequency interpretations. A clear prediction is that $\Delta\nu$
should decrease not only when $\nu_2$ increases (as observed) but also
when it sufficiently decreases (Fig.\,\ref{fig:rpm}).

For a precise match between model and observations, additional free
parameters are required. Stella \& Vietri (1999) propose that the
orbital eccentricity $e$ systematically varies with orbital
frequency. A critical discussion of the degree to which the precession
model and the beat-frequency model can each fit the data can be found
in Psaltis et al. (1999b). Vietri \& Stella (1998) and Armitage et
al. (1999) have performed calculations relevant to the problem of
sustaining the tilted orbits required for Lense-Thirring precession in
a viscous disk, where the Bardeen-Petterson effect (1975) drives the
matter to the orbital plane. Karas (1999a,b) calculated frequencies
and light curves produced by clumps orbiting the neutron star in
orbits similar to the ones discussed here. Miller (1999d) calculated
the effects of radiation forces on Lense-Thirring precession. Kalogera
\& Psaltis (1999) explored how the Lense-Thirring precession
interpretation constrains neutron star structure.

Relatively high neutron star masses (1.8--2\msun), relatively stiff
equations of state, and neutron star spin frequencies in the
300--900\,Hz range follow from this model. Because it
requires no neutron star, but only a relativistic accretion disk to
work, the model can also be applied to black holes. Stella et
al. (1999b) propose it explains the frequency correlations discussed in
\S\ref{sect:slowqpo} (Fig.\,\ref{fig:rpm}).

The idea that the three most prominent frequencies observed are in
fact the three main general-relativistic frequencies characterizing a
free-particle orbit is fascinating. However, questions remain. How
precessing and eccentric orbits can survive in a disk is not a priori
clear (see Psaltis \& Norman 1999 for a possible way to obtain these
frequencies from a disk). How the flux is modulated at the predicted
frequencies, why the basic, orbital, frequency $\nu_2$ varies with
luminosity, and how the burst oscillations fit in are all open
questions. With respect to the burst oscillations the model requires
other explanations than neutron star spin (\S\ref{sect:burstosc}) and
these are being explored (Stella 1999, Klu\'zniak 1999, Psaltis \&
Norman 1999).

\subsection{Photon bubble model}\label{sect:pbm}

A model based on numerical radiation hydrodynamics was proposed by
Klein et al. (1996b) for the kHz QPOs in Sco\,X-1. In this model
accretion takes place by way of a magnetic funnel within which
accretion is super-Eddington, so that photon bubbles form which rise
up by buoyancy and burst at the top in quasi-periodic sequence. In
some of the simulations one or two strong QPO peaks are found, whose
frequencies increase with accretion rate, as observed (R. Klein,
priv. comm.). The model stands out by not requiring rotational
phenomena to explain the QPOs, and does not naturally produce beat
frequencies. In recent work, attention with respect to this model has
shifted to the classical accreting pulsars for which it was originally
conceived (Klein et al. 1996a; \S\ref{sect:cenx3}).

\subsection{Disk mode models}\label{sect:disk}

In several models the observed frequencies are identified with
oscillation modes of an accretion disk. From an empirical point of
view, these models fall into two classes. Some can be seen as
``implementations'' of a beat-frequency (e.g. Alpar et al 1997) or a
precession (Psaltis \& Norman 1999) model that they provide ways in
accretion disk physics to produce signals at the frequencies occurring
in those models. Others produce new frequencies. In all disk models
one of the two kHz QPOs is a Keplerian orbital frequency at some
radius in the disk.

The disk transition layer models (Titarchuk \& Muslimov 1997,
Titarchuk et al. 1998,1999, Osherovich \& Titarchuk 1999a,b,
Titarchuk \& Osherovich 1999) have evolved into a description (the
``two-oscillator model'') where $\nu_1$ is identified with the
Keplerian frequency at the outer edge of a viscous transition layer
between Keplerian disk and neutron-star surface.  Oscillations in this
layer occur at two low frequencies, producing the noise break and a
low frequency QPO (\S\ref{sect:slowqpo}).  Additionally, blobs
described as being thrown out of this layer into a magnetosphere
oscillate both radially and perpendicular to the disk, producing two
harmonics of another low-frequency QPO (in the Z sources this is the
HBO) as well as the upper kHz peak. So, altogether this description
provides six frequencies which can all fit observed frequencies.
Further work on disk oscillations was performed by Lai (1998, 1999),
Lai et al. (1999) and Ghosh (1998; see also Moderski \& Czerny 1999).

\section{Final remark}

RXTE has opened up a window that allows us to see down to the very
bottoms of the potential wells of some neutron stars and perhaps to
near to the horizons of some black holes.  Three millisecond phenomena
have been found, whose interpretation relies explicitly on our
description of strong-field gravity and neutron-star structure. In
order to take full advantage of this, it will be necessary to observe
the new phenomena with larger instruments (in the 10\,m$^2$
class). This will allow to follow the motion of clumps of matter
orbiting in strong gravity and of hot spots corotating on neutron star
surfaces and thereby to to map out curved spacetime near accreting
compact objects and measure parameters such as the compactness of
neutron stars and the spin of black holes.

{\bf Acknowledgements:} It is a pleasure to acknowledge the help of
many colleagues who either made data available before publication,
sent originals of figures, read versions of the manuscript or provided
insightful discussion: Didier Barret, Lars Bildsten, Deepto
Chakrabarty, Wei Cui, Eric Ford, Peter Jonker, Richard Klein, Fred
Lamb, Phil Kaaret, Craig Markwardt, Mariano M\'endez, Cole Miller,
Mike Nowak, Dimitrios Psaltis, Luigi Stella, Tod Strohmayer, Rudy
Wijnands and Will Zhang. This work was supported in part by the
Netherlands Organization for Scientific Research (NWO) and the
Netherlands Research School for Astronomy (NOVA).

\def\aj{{AJ}}                   
\def\araa{{\it ARA\&A\ }}             
\def\apj{{\it ApJ\ }}                 
\def\apjl{{\it ApJ\ }}                
\def\apjs{{\it ApJS\ }}               
\def\apss{{Ap\&SS}}             
\def\aap{{\it A\&A\ }}                
\def\aapr{{\it A\&A~Rev.}}          
\def\aaps{{\it A\&AS}}              
\def\azh{{AZh}}                 
\def\baas{{BAAS}}               
\def\jrasc{{JRASC}}             
\def\memras{{MmRAS}}            
\def\mnras{{\it MNRAS\ }}             
\def\pra{{Phys.~Rev.~A}}        
\def\prb{{Phys.~Rev.~B}}        
\def\prc{{\it Phys.~Rev.~C\ }}        
\def\prd{{\it Phys.~Rev.~D\ }}        
\def\pre{{Phys.~Rev.~E}}        
\def\prl{{\it Phys.~Rev.~Lett.\ }}    
\def\pasp{{PASP}}               
\def\pasj{{\it PASJ\ }}               
\def\qjras{{QJRAS}}             
\def\skytel{{S\&T}}             
\def\solphys{{Sol.~Phys.}}      
\def\sovast{{\it Soviet~Ast.\ }}      
\def\ssr{{\it Space~Sci.~Rev.\ }}     
\def\zap{{ZAp}}                 
\def\nat{{\it Nature\ }}              
\def\iaucirc{{\it IAU~Circ. No.}}       
\def\aplett{{Astrophys.~Lett.}} 
\def\apspr{{Astrophys.~Space~Phys.~Res.}}
\def\bain{{Bull.~Astron.~Inst.~Netherlands}} 
\def\fcp{{Fund.~Cosmic~Phys.}}  
\def\gca{{Geochim.~Cosmochim.~Acta}}   
\def\grl{{Geophys.~Res.~Lett.}} 
\def\jcp{{J.~Chem.~Phys.}}      
\def\jgr{{J.~Geophys.~Res.}}    
\def\jqsrt{{J.~Quant.~Spec.~Radiat.~Transf.}}
\def\memsai{{Mem.~Soc.~Astron.~Italiana}}
\def\nphysa{{Nucl.~Phys.~A}}   
\def\nphysb{{\it Nucl.~Phys.~B\ }}   
\def\physrep{{Phys.~Rep.}}   
\def\physscr{{Phys.~Scr}}   
\def\planss{{Planet.~Space~Sci.}}   
\def\procspie{{Proc.~SPIE}}   

\parindent=0pt
\section*{Literature cited}\def\xxx{}

\end{document}